\documentclass[]{article}
\date{}
\usepackage{graphicx}
\usepackage{float}
\usepackage{amsmath}
\usepackage{amssymb}
\usepackage[title]{appendix}
\newfloat{EqFloat}{h}{EqFloatFile}
\usepackage[top=1in, bottom=1in, left=1in, right=1in]{geometry}
\usepackage[title]{appendix}
\usepackage{hyperref}
\usepackage{setspace}
\begin{document}
\title{Capacity Analysis of LTE-Advanced HetNets with Reduced Power Subframes and Range Expansion}
\author{Arvind~Merwaday$^\natural$, Sayandev~Mukherjee$^\sharp$, and~Ismail~G\"uven\c{c}$^\natural$\\
Email: $^\natural${\tt \{amerw001, iguvenc\}@fiu.edu}, $^\sharp${\tt smukherjee@docomoinnovations.com}}
\maketitle

\begin{abstract}
The time domain inter-cell interference coordination techniques specified in LTE Rel. 10 standard improves the throughput of picocell-edge users by protecting them from macrocell interference. On the other hand, it also degrades the aggregate capacity in macrocell because the macro base station (MBS) does not transmit data during certain subframes known as almost blank subframes. The MBS data transmission using reduced power subframes was standardized in LTE Rel. 11, which can improve the capacity in macrocell while not causing high interference to the nearby picocells. In order to get maximum benefit from the reduced power subframes, setting the key system parameters, such as the amount of power reduction, carries critical importance. Using stochastic geometry, this paper lays down a theoretical foundation for the performance evaluation of heterogeneous networks with reduced power subframes and range expansion bias. The analytic expressions for average capacity and 5th percentile throughput are derived as a function of transmit powers, node densities, and interference coordination parameters in a heterogeneous network scenario, and are validated through Monte Carlo simulations. Joint optimization of range expansion bias, power reduction factor, scheduling thresholds, and duty cycle of reduced power subframes are performed to study the trade-offs between aggregate capacity of a cell and fairness among the users. To validate our analysis, we also compare the stochastic geometry based theoretical results with the real MBS deployment (in the city of London) and the hexagonal-grid model. Our analysis shows that with optimum parameter settings, the LTE Rel. 11 with reduced power subframes can provide substantially better performance than the LTE Rel. 10 with almost blank subframes, in terms of both aggregate capacity and fairness.
\end{abstract}

\noindent{\bf keywords: fairness, FeICIC, HetNets, LTE-Advanced, performance analysis, Poisson point process, PPP, reduced power ABS, reduced power subframes.}

\section{Introduction}
Cellular networks are witnessing an exponentially increasing data traffic from mobile users. Heterogeneous networks (HetNets) offer a promising way of meeting these demands. They are composed of small-size cells such as micro-, pico-, and femto-cells overlaid on the existing macrocells to increase the frequency reuse and capacity of the network. Since the base stations (BSs) of different tiers use different transmission powers and typically a frequency reuse factor of one, analyzing and mitigating the interference at an arbitrary user equipment (UE) is a challenging task.

\subsection{Related Work on Evaluation Methodology}
Different approaches have been used in the literature for the performance evaluation of HetNets. The traditional simulation models with BSs placed on a hexagonal grid are highly idealized and may typically require complex and time-consuming system-level simulations. On the other hand, models based on stochastic geometry and spatial point processes provide a tractable and computationally efficient alternative for performance evaluation of HetNets \cite{6524460}-\nocite{6042301, 5962727}\cite{5621983}. \emph{Poisson point process} (PPP) based models have been recently used extensively in the literature for performance evaluation of HetNets. However, as the macro base station (MBS) locations are carefully planned during the deployment process, PPP based models may not be viable for capturing real MBS locations, due to some points of the process being very close to each other. \emph{Matern hardcore point process} (HCPP) provides a more accurate alternative spatial model for MBS locations. In HCPPs, the distance between any two points of the process is greater than a minimum distance predefined by hard core parameter. HCPP models are relatively more complicated due to the non existence of the probability generating functional \cite{6524460}. Also, HCPP has a flaw of underestimating the intensity of the points that can coexist for a given hard core parameter \cite{5934671}. Hence, HCPP models are not as tractable and simple as the PPP models.

With PPPs, using simplifying assumptions, such as Rayleigh fading channel model, and a path-loss exponent of four, we can obtain closed form expressions for aggregate interference and outage probability. Therefore, use of PPP models for performance evaluation of HetNets is appealing due to their simplicity and tractability \cite{5226957}. Furthermore, the PPP based models provide reasonably close performance results when compared with the real BS deployments. In particular, results in \cite{5962727} show that, when compared with real BS deployments, PPP and hexagonal grid based models for BS locations provide a lower bound and an upper bound, respectively, on the outage probabilities of UEs. Also, the PPP based models are expected to provide a better fit for analyzing denser HetNet deployments due to higher degree of randomness in small-cell deployments \cite{6042301}. In this paper, due to their simplicity and reasonable accuracy, we will use PPP based models to characterize and understand the behavior of HetNets in terms of various design parameters.

\subsection{Use of PPP Based Models for LTE-Advanced HetNet Performance Evaluation}
The existing literature has numerous papers based on the PPP model for analyzing HetNets. Using PPPs, the basic performance indicators such as coverage probability and average rate of a UE are analyzed in \cite{sayan-icc-ws}-\nocite{Robert_TSP_2012, Andrews_Globecom_2011_ICIC}\cite{dhillon-ita}. The use of range expansion bias (REB) in the picocell enables it to associate with more UEs and thereby improves the offloading of UEs to the picocells. The effect of REB on the coverage probability is studied in \cite{Sayan_ICC_2012, Andrews_Globecom2011_RE}. However, with range expansion, the offloaded UEs at the edge of picocells experience high interference from the macrocell. This necessitates a coordination mechanism between the MBSs and pico base stations (PBSs) to protect the picocell-edge UEs from the MBS interference. While \cite{6042301, 5962727, RealBS_Lee} considers a homogeneous cellular network, \cite{Andrews_Globecom2011_RE} considers a HetNet with range expansion. The authors of \cite{6042301, 5962727, Andrews_Globecom2011_RE} have obtained the information of real BS locations in an urban area from a cellular service provider. On the other hand, the authors of \cite{RealBS_Lee} have obtained the BS location information from an open source project \cite{opencellid} that provides approximate locations of the BSs around the world.

To mitigate the interference problems in HetNets, different \emph{enhanced inter-cell interference coordination} (eICIC) techniques have been specified in LTE Rel. 10 of 3GPP which includes time-domain, frequency domain and power control techniques \cite{Lopez_perez2011HetNet}. In the time domain eICIC technique, MBS transmissions are muted during certain subframes and no data is transmitted to macro UEs (MUEs). The picocell-edge users are served by PBS during these subframes (coordinated subframes) and thereby protecting the picocell-edge users from MBS interference. The eICIC technique using REB is studied well in the literature by analyzing its effects on the rate coverage \cite{Jeff_2013_arxiv, Jeff_IEEE_Trans_2013} and on the average per-user capacity \cite{Asilomar_2011,CapAnalysis_Globecom_2013}. However, in the simulations of \cite{R1-113806_Panasonic}, the MBS transmits at reduced power (instead of muting the MBS completely) during the coordinated subframes (CSFs) to serve only its nearby UEs. Therein, the use of reduced power subframes during CSFs is shown to improve the HetNet performance considerably in terms of the trade-off between the cell-edge and average throughputs. Later on, \emph{reduced power subframe} transmission have also been standardized under LTE Rel.~11 of 3GPP, and commonly referred therein as \emph{further-enhanced ICIC} (FeICIC). In another study \cite{Morimoto_IEICE_2013}, simulation results show that the FeICIC is less sensitive to the duty-cycle of CSFs than the eICIC. In \cite{6289196}, 3GPP simulations are used to study and compare the eICIC and FeICIC techniques for different REBs and \emph{almost blank subframe} densities. Therein, the amount of power reduction in the reduced power subframes is made equivalent to REB and its optimality is not justified.

\subsection{Contributions}
In authors' earlier work, analytic expressions using PPPs for coverage probability of an arbitrary UE is derived in \cite{sayan-icc-ws} which has been extended to spectral efficiency (SE) derivations in \cite{Asilomar_2011,CapAnalysis_Globecom_2013} by considering eICIC and range expansion. Reduced power subframes, which are standardized in LTE Rel. 11 \cite{LTE_Rel_11_Overview}, are not analytically studied in literature to our best knowledge.

In the present work, generalized SE expressions are derived considering the FeICIC which includes eICIC and no eICIC as the two special cases. In this analytic framework that uses reduced power subframes and range expansion, expressions for the average SE of UEs and the 5th percentile throughput are derived. These expressions are validated through Monte Carlo simulations. Details of the simulation model are documented explicitly, and the Matlab codes can be accessed through \cite{MpactWebsite} for regenerating the results. The optimization of key system parameters is analyzed with a perspective of maximizing both aggregate capacity in a cell and the proportional fairness among its users. Using these results, insights are developed on the configuration of FeICIC parameters, such as the power reduction level, range expansion bias, duty cycle of CSFs, and scheduling thresholds. The 5th and 50th percentile capacities are also analyzed to determine the trade-offs associated with FeICIC parameter adaptation. Further, we compare the 5th percentile SE results from PPP model with the real MBS deployment \cite{SiteFinder} and the hexagonal grid model.

\section{System Model}
We consider a two-tier HetNet system with MBS, PBS and UE locations modeled as two-dimensional homogeneous PPPs of intensities $\lambda$, $\lambda'$ and $\lambda_{\rm u}$, respectively. Both the MBSs and the PBSs share a common transmission bandwidth. The MBSs employ reduced power subframes, in which they transmit at reduced power levels to prevent high interference to the picocell UEs (PUEs). On the other hand, the PBSs transmit at full power during all the subframes.

\begin{figure}[h!]
\center
\includegraphics[width=5in]{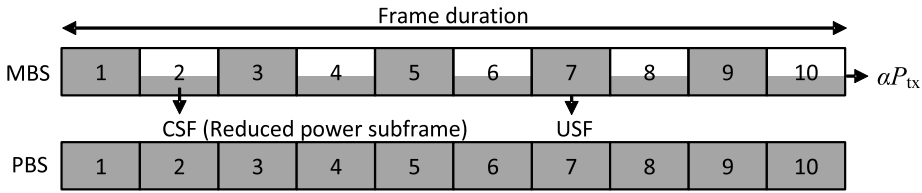}
\vspace{-3mm}
\caption{Frame structure with reduced power subframes, transmitted with a duty cycle of $\beta=0.5$.}
\label{fig:FrameStructure}
\end{figure}
The frame structure with reduced power subframes is shown in Figure~\ref{fig:FrameStructure}. During uncoordinated subframes (USFs), the MBS transmits data and control signals at full power $P_{\rm tx}$ and during CSFs, it transmits at a reduced power $\alpha P_{\rm tx}$, where $0\leq\alpha\leq 1$ is the power reduction factor. The PBS transmits the data, control signals and \emph{cell reference symbol} with power $P'_{\rm tx}$ during all the subframes. Setting $\alpha = 0$ corresponds to eICIC; and $\alpha = 1$ corresponds to no eICIC case. 

Define $\beta$ as the duty cycle of USFs, i.e., ratio of the number of USFs to the total number of sub-frames in a frame. Then, $(1-\beta)$ is the duty cycle of CSF/reduced power subframes. Let $K$ and $K'$ be the factors that account for geometrical parameters such as the transmitter and receiver antenna heights of the MBS and the PBS, respectively. Then, the effective transmitted powers of MBS during USFs is $P=P_{\rm tx}K$, MBS during CSFs is $\alpha P$, and PBS during USF/CSF is $P'=P'_{\rm tx}K'$. For an arbitrary UE, let the nearest MBS at a distance $r$ be its macrocell of interest (MOI) and the nearest PBS at a distance $r'$ be its picocell of interest (POI). Then, assuming Rayleigh fading channel, the \emph{reference symbol received power} from the MOI and the POI are given by,
\begin{align}
S(r) = \frac{PH}{r^\delta} \label{rsrp}, \ \ \ \ \ 
S'(r') = \frac{P'H'}{(r')^\delta},
\end{align}
respectively, where the random variables $H \sim \mbox{Exp}(1)$ and $H' \sim \mbox{Exp}(1)$ account for Rayleigh fading. Define an interference term, $Z$, as the total interference power at a UE during USFs from all the MBSs and the PBSs, excluding the MOI and the POI. Similarly, define $Z'$ as the total interference power during CSFs. We assume that there is no frame synchronization across the MBSs and therefore irrespective of whether the MOI is transmitting a USF or a CSF, the interference at UE has the same distribution in both cases, and is independent of both $S(r)$ and $S'(r')$. Then, an arbitrary UE experiences the following four SIRs:
\begin{align}
\Gamma=&\frac{S(r)}{S'(r')+Z}, \rightarrow \mbox{USF SIR from MOI} \label{sirmacusf}\\
\Gamma' =&\frac{S'(r')}{S(r)+Z}, \rightarrow \mbox{USF SIR from POI} \label{sirpicusf}\\
\Gamma_{\rm CSF}=&\frac{\alpha S(r)}{S'(r')+Z}, \rightarrow \mbox{CSF SIR from MOI} \label{sirmaccsf}\\
\Gamma'_{\rm CSF}=& \frac{S'(r')}{\alpha S(r)+Z}. \rightarrow \mbox{CSF SIR from POI} \label{sirpiccsf}
\end{align}

\subsection{UE Association}
In \eqref{sirmaccsf} and \eqref{sirpiccsf}, it can be noted that $\Gamma_{\rm csf}$ and $\Gamma'_{\rm csf}$ are directly affected by $\alpha$ and hence their usage will make the cell selection process dependent on $\alpha$. Thus, we consider $\Gamma$ and $\Gamma'$ to minimize the dependence of the cell selection process on $\alpha$.

\begin{figure}[h!]
\center
\includegraphics[width=6in]{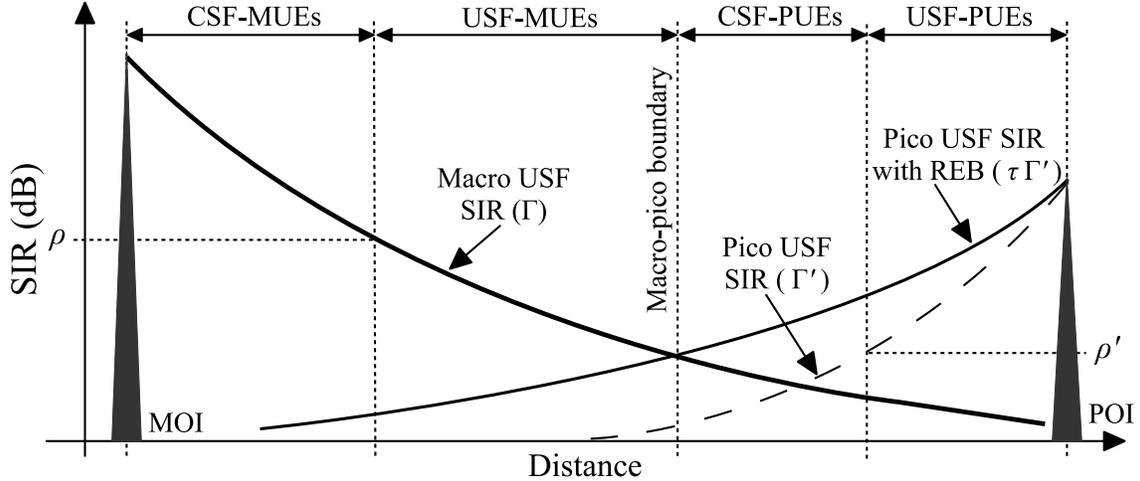}
\vspace{-3mm}
\caption{Illustration of UE association criteria.}
\label{fig:SIR}
\end{figure}
The cell selection process using $\Gamma$, $\Gamma'$ and the REB $\tau$ can be explained with reference to Figure~\ref{fig:SIR}. If $\tau\Gamma'$ is less than $\Gamma$, then the UE is associated with the MOI, otherwise with the POI. After the cell selection, the UE is scheduled either in USF or in CSF based on the scheduling thresholds $\rho$ (for MUE) and $\rho'$ (for PUE). In macrocell, if $\Gamma$ is less than $\rho$ then the UE is scheduled to USF, otherwise to CSF. Similarly, in picocell, if $\Gamma'$ is greater than $\rho'$ then the UE is scheduled to USF, otherwise to CSF (to protect it from macrocell interference). The cell selection and scheduling conditions can be combined and formulated as:
\begin{align}
\mbox{If } \Gamma>\tau\Gamma' \mbox{ and } \Gamma\leq\rho & \rightarrow \mbox{USF-MUE}, \label{usf_mue}\\
\mbox{If } \Gamma>\tau\Gamma' \mbox{ and } \Gamma>\rho & \rightarrow \mbox{CSF-MUE},\label{csf_mue}\\
\mbox{If } \Gamma\leq\tau\Gamma' \mbox{ and } \Gamma' > \rho' & \rightarrow \mbox{USF-PUE},\label{usf_pue}\\
\mbox{If } \Gamma\leq\tau\Gamma' \mbox{ and } \Gamma' \leq \rho' & \rightarrow \mbox{CSF-PUE}. \label{csf_pue}
\end{align}

\begin{figure}[h!]
\vspace{-7mm}
\center
\includegraphics[width=4.5in]{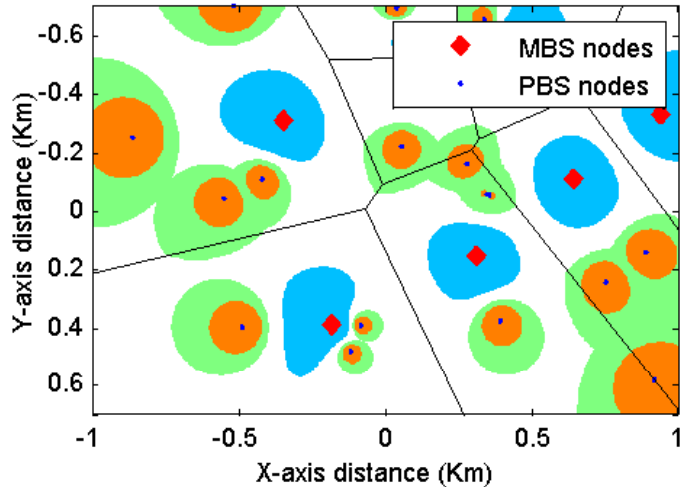}
\vspace{-5mm}
\caption{Illustration of two-tier HetNet layout. In picocells, the coverage regions for USF- and CSF-PUEs are colored in orange and green, respectively. Whereas in macrocells, the coverage regions for USF- and CSF-MUEs are colored in white and blue, respectively.}
\label{fig:Coverage_Area}
\end{figure}
A sample layout of MBSs and PBSs with their coverage areas for the four different UE categories are illustrated in Figure~\ref{fig:Coverage_Area}. Note that in the related work of \cite{Jeff_2013_arxiv}, the UE association criteria are based on the average reference symbol received power at UE, where as our model is based on the SIR at UE, it also encompasses the FeICIC mechanism. In \cite{Jeff_2013_arxiv}, the boundary between the USF-PUEs (picocell area) and the CSF-PUEs (range expanded area) is fixed due to the fixed transmit power of PBS. On the other hand, in our approach, the boundary between USF and CSF users can be controlled using $\rho$ in macrocell and $\rho'$ in picocell, the parameters which play an important role during optimization as will be shown in Section~\ref{sec:optimization}.

Using \eqref{rsrp}-\eqref{sirpiccsf}, it can be shown that the two SIRs $\Gamma_{\rm CSF}$ and $\Gamma'_{\rm CSF}$ could be expressed in terms of $\Gamma$ and $\Gamma'$ as,
\begin{align}
\Gamma_{\rm csf} = \alpha\Gamma,\ \ \ \ \Gamma'_{\rm csf} = \frac{\Gamma'(1+ \Gamma)}{1+\Gamma[\alpha(\Gamma'+1)-\Gamma']}.
\label{SIR_picusf1}
\end{align}
Hence, knowing the statistics of $\Gamma$ and $\Gamma'$, particularly their \emph{joint probability density function} (JPDF), would provide a complete picture of the SIR statistics of the HetNet system. We first derive an expression for \emph{joint complementary cumulative distribution function} (JCCDF) of $\Gamma$ and $\Gamma'$ in Section~\ref{sec:JCCDF}. Then we differentiate the JCCDF with respect to $\gamma$ and $\gamma'$ to get the expression for JPDF in Section~\ref{sec:JPDF}, which will then be used for spectral efficiency analysis.

\section{Derivation of Joint SINR Distribution}
\subsection{JCCDF of $\Gamma$ and $\Gamma'$}
\label{sec:JCCDF}
From \eqref{rsrp}, we know that $S(r)$ and $S'(r')$ are exponentially distributed with mean $P/r^\delta$ and $P'/(r')^\delta$, respectively. For brevity, substitute $S(r) = X$ and $S'(r') = Y$ in \eqref{sirmacusf} and \eqref{sirpicusf}:
\begin{align}
\Gamma=\frac{X}{Y+Z}, \ \ \ \ \Gamma'=\frac{Y}{X+Z}. \label{temp1}
\end{align}
Using \eqref{temp1} it can be easily shown that the product $\Gamma\Gamma'$ has a maximum value of 1.

\begin{figure}[h!]
\center
\includegraphics[width=4.5in]{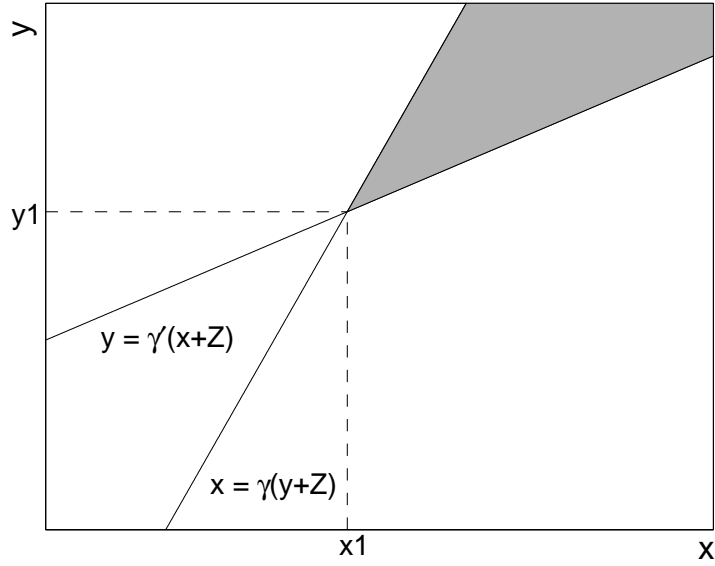}
\vspace{-6mm}
\caption{Illustration of the integration region in the JPDF of $X$ and $Y$. Shaded region indicates the integration region in order to compute the JCCDF.}
\label{fig:JCCDFIntRegion}
\end{figure}
Let, $R$ and $R'$ be the random variables denoting the distances of MOI and POI from a UE. Then, the JCCDF of $\Gamma$ and $\Gamma'$ conditioned on $R=r, R'=r'$ is given by,
\begin{align}
\mathbb{P}\{\Gamma > \gamma,\Gamma' > \gamma' \bigm | R=r, R'=r'\} &= \mathbb{E}_Z\bigg[\mathbb{P}\big\{X>\gamma(Y+Z), Y>\gamma'(X+Z)\big\}\bigg], \nonumber \\
&= \mathbb{E}_Z \left[ \int_{y1}^{+\infty}f_Y(y) \int_{\gamma(y+Z)}^{y/\gamma'-Z} f_X(x) \,{\rm d}x\,{\rm d}y \right], \label{JCCDF_int}
\end{align}
for $\gamma>0, \gamma'>0, \gamma \gamma' < 1$. Here, $f_{\rm X}(x)=\frac{r^\delta}{P}\exp\left({-\frac{r^\delta}{P}x}\right)$, $f_{\rm Y}(y)=\frac{(r')^\delta}{P'}\exp\left({-\frac{(r')^\delta}{P'}y}\right)$, and the integration limit $y_1 = \gamma' Z \left( \frac{1+\gamma}{1-\gamma \gamma'} \right).$ The integration region of \eqref{JCCDF_int} is graphically represented in Figure~\ref{fig:JCCDFIntRegion}. By solving the integration as shown in Appendix~\ref{App:JCCDFDerive}, we can obtain a closed form expression for the conditional JCCDF as
\begin{align}
\mathbb{P}\{\Gamma > \gamma,\Gamma' > \gamma'| R=r, R'=r'\} = \frac{(1-\gamma\gamma')\mathcal{L}_Z \left(\frac{1}{1-\gamma\gamma'} \left( \frac{\gamma(1+\gamma')r^\delta}{P}+\frac{\gamma'(1+\gamma)(r')^\delta}{P'}\right)\right)}{\left[ 1+\gamma\frac{P'}{P} \left(\frac{r}{r'}\right)^\delta\right] \left[ 1+\gamma'\frac{P}{P'} \left(\frac{r'}{r}\right)^\delta\right]},
\label{eq:JCCDF}
\end{align}
for $\gamma>0, \gamma'>0,$ and $\gamma \gamma' < 1$, where $\mathcal{L}_Z(s)$ is the Laplace transform of the total interference $Z$.

Expression for $\mathcal{L}_{\rm Z}(s)$ can be derived as follows. We assume that the interfering MBSs of a UE are frame asynchronous and subframe synchronous. Moreover, locations of the USFs and CSFs are uniformly randomly distributed, with a USF duty cycle of $\beta$ for all the MBSs. Hence, each interfering MBS transmits USFs with probability $\beta$ and CSFs with probability $(1-\beta)$ and the tier of MBSs can be split into two tiers, one tier of MBSs transmitting only USFs and other transmitting only CSFs. These two tiers are independent PPPs with intensities $\lambda\beta$ and $\lambda(1-\beta)$. Therefore, the FeICIC scenario can be modeled using three independent PPPs as illustrated in Table~\ref{tab:PPPs}.
\begin{table}[h!]
\caption{PPP parameters for USF MBSs, CSF MBSs, and PBSs.}
\label{tab:PPPs}
\center
\begin{tabular}{|l|l|l|p{0.9cm}|p{2.5cm}|}
\hline
 {\bf BS type} & {\bf PPP} & {\bf Intensity} & {\bf Tx. power} & {\bf Distance of UE to nearest BS} \\ \hline
USF-MBSs & $\Phi_{\rm USF}$ & $\beta\lambda$ & $P$ & $r$ \\ \hline
CSF-MBSs & $\Phi_{\rm CSF}$ & $(1-\beta)\lambda$ & $\alpha P$ & $r$ \\ \hline
PBSs & $\Phi'$ & $\lambda'$ & $P'$ & $r'$ \\ \hline
\end{tabular}
\end{table}

Let, $I_{\rm USF}(r)$, $I_{\rm CSF}(r)$, and $I'(r')$ be the interference at UE from all interfering USF-MBSs, CSF-MBSs and PBSs. Then, the total interference is $Z = I_{\rm USF}(r)+I_{\rm CSF}(r)+I'(r')$. Using \cite[Corollary~1]{Sayan_ICC_2011}, parameters in Table~\ref{tab:PPPs}, and assuming $\delta=4$, we can derive the Laplace transform of $Z$ in \eqref{eq:JCCDF} to be,
\begin{align}
\mathcal{L}_{\rm Z}(s) =& \exp\bigg\{-\pi\beta\lambda\sqrt{Ps}\ \left[\frac{\pi}{2}-\tan^{-1}\left(\frac{r^2}{\sqrt{Ps}}\right)\right] - \pi(1-\beta)\lambda\sqrt{\alpha Ps}\ \left[\frac{\pi}{2}-\tan^{-1}\left(\frac{r^2}{\sqrt{\alpha Ps}}\right)\right] \nonumber\\
& \ \ \ \ \ \ \ \ - \pi\lambda'\sqrt{P's}\ \left[\frac{\pi}{2} - \tan^{-1}\left(\frac{(r')^2}{\sqrt{P's}}\right)\right]\bigg\}. \label{Lz}
\end{align}

\subsection{JPDF of $\Gamma$ and $\Gamma'$}
\label{sec:JPDF}
The conditional JPDF of $\Gamma$ and $\Gamma'$,
\begin{align}
f_{\Gamma,\Gamma'\bigm|R,R'}(\gamma,\gamma'\bigm|r,r')= \mathbb{P}\{\Gamma = \gamma,\Gamma' = \gamma'| R=r, R'=r'\}
\end{align}
can be derived by differentiating the JCCDF in \eqref{eq:JCCDF} with respect to $\gamma$ and $\gamma'$. Detailed derivation of conditional probability JPDF is provided in Appendix~\ref{App:JPDFDerive}. Using the theorem of conditional probability we can write
\begin{align}
f_{\Gamma,\Gamma',R,R'}(\gamma,\gamma',r,r')= f_{\Gamma,\Gamma'\bigm|R,R'}(\gamma,\gamma'\bigm|r,r') f_R(r) f_{R'}(r'),
\end{align}
where, the PDFs of $R$ and $R'$ are $f_R(r) = 2\pi\lambda r e^{-\lambda\pi r^2}$ and $f_{R'}(r') = 2\pi\lambda' r' e^{-\lambda'\pi(r')^2}$, respectively. We can then express the unconditional JPDF of $\Gamma$ and $\Gamma'$ as,
\begin{align}
f_{\Gamma,\Gamma'}(\gamma,\gamma') &= \int_{d_{\rm min}}^\infty \int_{d'_{\rm min}}^\infty f_{\Gamma,\Gamma',R,R'}(\gamma,\gamma',r,r') \,{\rm d}r' \,{\rm d}r\nonumber\\
&= \int_{d_{\rm min}}^\infty \int_{d'_{\rm min}}^\infty \!\!\!\!\!\!f_{\Gamma,\Gamma'\bigm|R,R'}(\gamma,\gamma'\bigm|r,r') f_R(r) f_{R'}(r') \,{\rm d}r' \,{\rm d}r, \label{eq:Jpdf_uncond}
\end{align}
where, we assume that a UE is served by a BS only if it satisfies the minimum distance constraints: UE should be located at distances of at least $d_{\rm min}$ from the MOI and $d'_{\rm min}$ from the POI.

\section{Spectral efficiency analysis}
In this section, the expressions for aggregate and per-user SEs categories are derived. Considering the JPDF of an arbitrary UE in \eqref{eq:Jpdf_uncond}, first the expressions for the probabilities that the UE belongs to each category are derived. Then, these expressions are used to derive the mean number of UEs of each category in a cell. These are followed by the derivation of the aggregate SE. Then, per-user SE expressions are obtained by dividing the aggregate SE by the mean number of UEs.

\subsection{MUE and PUE Probabilities}
\begin{figure}[h!]
\center
\includegraphics[width=4.5in]{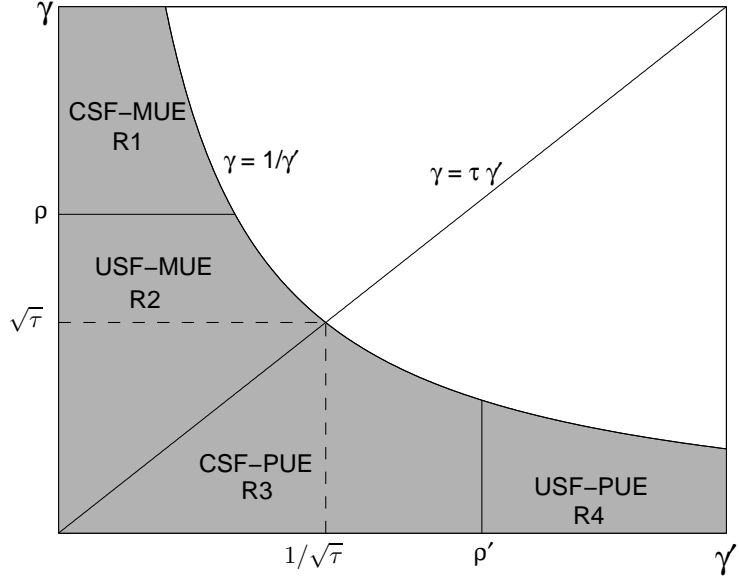}
\vspace{-6mm}
\caption{Illustration of the integration regions in the JPDF of $\Gamma$ and $\Gamma'$. Shaded regions indicate the integration regions to compute the probabilities of a UE belonging to different categories.}
\label{fig:Capacity_int_regions}
\end{figure}

Depending on the SIRs $\Gamma$ and $\Gamma'$, a UE can be one of the four types: USF-MUE, CSF-MUE, USF-PUE or CSF-PUE. Given that the UE is located at a distance $r$ from its MOI and $r'$ from its POI, probabilities of the UE belonging to each type can be found by integrating the conditional JPDF over the regions whose boundaries are set by the cell selection conditions in \eqref{usf_mue}-\eqref{csf_pue}. Based on these conditions the integration regions for different UE categories are shown in Figure~\ref{fig:Capacity_int_regions}.

The probability that a UE is a CSF-MUE can be found by integrating the JPDF over the region R1,
\begin{align}
P_{\rm csf} =& \mathbb{P}\{\Gamma>\tau\Gamma',\Gamma>\rho\} = \int_{\rho}^{\infty}\!\!\!\!\int_{0}^{\min\left(\frac{1}{\gamma},\frac{\gamma}{\tau}\right)} f_{\Gamma, \Gamma'}(\gamma,\gamma') \,{\rm d}\gamma' \,{\rm d}\gamma. \label{muecsf_condprob}
\end{align}
To form concise equations, let us define an integral function,
\begin{align}
G(g,{\rm R}i) = \int\int_{{\rm R}i} g(\gamma,\gamma') f_{\Gamma,\Gamma'}(\gamma,\gamma') \,\mathrm{d}\gamma'\,\mathrm{d}\gamma,
\end{align}
where, $g$ is a function of $\gamma$ and $\gamma'$, $\rm R\it i$ for $i=1,2,3,4$ is the integration region as defined in Figure~\ref{fig:Capacity_int_regions}. Then, \eqref{muecsf_condprob} can be written as,
\begin{align}
P_{\rm csf} = \mathbb{P}\{\Gamma>\tau\Gamma',\Gamma>\rho\} = G(1,\mathrm{R1}).\label{Pcsfmue}
\end{align}
Similarly, the conditional probabilities that a UE is a USF-MUE, USF-PUE or CSF-PUE are respectively given as,
\begin{align}
P_{\rm usf} =& \mathbb{P}\{\Gamma>\tau\Gamma',\Gamma\leq\rho\} = G(1,\mathrm{R2}),\label{Pusfmue}\\
P'_{\rm usf} =& \mathbb{P}\{\Gamma\leq\tau\Gamma',\Gamma'\geq \rho'\} = G(1,\mathrm{R4}),\\
P'_{\rm csf} =& \mathbb{P}\{\Gamma\leq\tau\Gamma',\Gamma' < \rho'\} = G(1,\mathrm{R3}). \label{Pcsfpue}
\end{align}

\subsection{Mean number of MUEs and PUEs}
Since the MBS locations are generated using PPPs, the coverage areas of all the MBSs resemble a Voronoi tessellation. Consider an arbitrary Voronoi cell. Let the
number of UEs in the cell be $N$, and number of CSF-MUEs in the cell be $M$. Then, $M$ is a random variable and the mean number of CSF-MUEs is given by,
\begin{align}
N_{\mathrm{csf}}&= E[M] = E\left[\sum_{n=1}^N 1\{\text{UE } n \text{ is a CSF-MUE}\}\right]\nonumber\\
&= E_N\left\{E\left[\sum_{n=1}^N 1\{\text{UE } n \text{ is a CSF-MUE}\} \bigm | N\right]\right\} \nonumber \\
&= E_N\left\{\sum_{n=1}^N E\bigg[1\{\text{UE } n \text{ is a CSF-MUE}\}\bigg]\right\},
\label{Ncsfmue}
\end{align}
where in~\eqref{Ncsfmue} we use the fact that the probability that any of the $N$ UEs in a cell being a CSF-MUE is independent of $N$. However, it is important to note that this is itself a consequence of our assumption that there is no limit on the number of CSF-MUEs per cell. Further, the event that any one of the UEs in a cell is a CSF-MUE is independent of the event that any other UE in that cell is a CSF-MUE, and all such events have the same probability of occurrence, namely $P_{\mathrm{csf}}$ given in~\eqref{Pcsfmue}. Then,
\begin{align}
N_{\mathrm{csf}} = E_N\left\{\sum_{n=1}^N P_{\mathrm{csf}}\right\} = E_N\left[N P_{\mathrm{csf}}\right] = P_{\mathrm{csf}}\,E[N].
\end{align}
Using~\cite[Lemma 1]{Yu_2012_TVT}, it can be shown that the mean number of UEs in a Voronoi cell is $\lambda_u/\lambda$. Therefore, the mean number of CSF-MUEs in a cell are given by,
\begin{align}
N_{\mathrm{csf}} = \frac{P_{\mathrm{csf}} \lambda_{\rm u}}{\lambda}.
\label{MeanNoCSFMUEs}
\end{align}
Similarly, the mean number of USF-MUEs, USF-PUEs and CSF-PUEs are respectively given by,
\begin{align}
N_{\mathrm{usf}} = \frac{P_{\mathrm{usf}} \lambda_{\rm u}}{\lambda},\ \ \ \ \ \ 
N'_{\mathrm{usf}} = \frac{P'_{\mathrm{usf}} \lambda_{\rm u}}{\lambda'},\ \ \ \ \ \ 
N'_{\mathrm{csf}} = \frac{P'_{\mathrm{csf}} \lambda_{\rm u}}{\lambda'}.
\end{align}

\subsection{Aggregate and Per-user Spectral Efficiencies}
We use Shannon capacity formula, $\log_2(1+SIR)$, to find the SE of each UE type. The mean aggregate SE of an arbitrarily located CSF-MUE can be found by
\begin{align}
C_{\rm csf}(\lambda,\lambda',\tau,\alpha,\rho,\beta) &= (1-\beta) \frac{\mathbb{E}\left[\log_2(1+\Gamma_{\rm CSF}) \bigm | \mbox{UE is a CSF-MUE}\right]}{P_{\rm csf}} \nonumber\\
&= (1-\beta) \frac{G(\log_2(1+\gamma_{\rm csf}),\rm R1)}{P_{\rm csf}},\nonumber\\
&= (1-\beta) \frac{G(\log_2(1+\alpha\gamma),\rm R1)}{P_{\rm csf}}.
\end{align}
Similarly, the mean aggregate SEs for USF-MUEs, USF-PUEs and CSF-PUEs can be respectively derived to be
\begin{align}
    C_{\text{usf}}(\lambda,\lambda',\tau,\alpha,\rho,\beta) =& \beta \frac{G(\log_2(1+\gamma),\rm R2)}{P_{\rm usf}},\label{eq:agg_SE_macro_usf} \\
    C'_{\text{usf}}(\lambda,\lambda',\tau,\alpha,\rho',\beta) =& \beta \frac{G(\log_2(1 + \gamma'),\rm R4)}{P'_{\rm usf}}, \label{eq:agg_SE_pico_usf}\\
    C'_{\text{csf}}(\lambda,\lambda',\tau,\alpha,\rho',\beta) =& (1-\beta) \frac{G(\log_2(1 + \gamma'_{\rm csf}),\rm R3)}{P'_{\rm csf}}, \label{eq:agg_SE_pico_csf}
\end{align}
where, $\gamma'_{\rm csf} = \frac{\gamma'(1+ \gamma)}{1+\gamma[\alpha(\gamma'+1)-\gamma']}$.
Then the corresponding per-user SEs are
\begin{align}
C_{\text{u,usf}}(\lambda,\lambda',\tau,\alpha,\rho,\beta) = & \frac{\lambda \ C_{\text{usf}}(\lambda,\lambda',\tau,\alpha,\rho,\beta)}{\lambda_{\rm u} \ P_{\rm usf}}, \label{eq:perUSF_MUE_SE}\\
C_{\text{u,csf}}(\lambda,\lambda',\tau,\alpha,\rho,\beta) = & \frac{\lambda \ C_{\text{csf}}(\lambda,\lambda',\tau,\alpha,\rho,\beta)}{\lambda_{\rm u} \ P_{\rm csf}}, \label{eq:perCSF_MUE_SE}\\
C'_{\text{u,usf}}(\lambda,\lambda',\tau,\alpha,\rho',\beta) = & \frac{\lambda' \ C'_{\text{usf}}(\lambda,\lambda',\tau,\alpha,\rho',\beta)}{\lambda_{\rm u} \ P'_{\rm usf}}, \label{eq:perUSF_PUE_SE} \\
C'_{\text{u,csf}}(\lambda,\lambda',\tau,\alpha,\rho',\beta) = & \frac{\lambda' \ C'_{\text{csf}}(\lambda,\lambda',\tau,\alpha,\rho',\beta)}{\lambda_{\rm u} \ P'_{\rm csf}}. \label{eq:perCSF_PUE_SE}
\end{align}

\subsection{5th Percentile Throughput}
\label{sec:5th_perc_thrpt}
The 5th percentile throughput reflects the throughput of cell-edge UEs. Typically the cell-edge UEs experience high interference and analyzing their throughput provides important information about the fairness among the users in a cell and the system performance.

Consider the JPDF expression in \eqref{eq:Jpdf_uncond}. The integration regions of the JPDF for different UE categories are shown in Figure~\ref{fig:Capacity_int_regions}. The SIR PDF of USF-MUEs can be evaluated by integrating the JPDF over $\gamma'$ in the region R2,
\begin{align}
f_{\Gamma}(\gamma) = \mathbb{P}\{\Gamma = \gamma \bigm | \mbox{UE is a USF-MUE}\} = \int_{0}^{\min\left(\frac{\gamma}{\tau}, \frac{1}{\gamma}\right)} f_{\Gamma,\Gamma'}(\gamma,\gamma') \,{\rm d}\gamma',
\end{align}
for $0 \leq \gamma \leq \rho$. The CDF expression can be derived as
\begin{align}
{\rm F}_{\Gamma}(\gamma_{\rm usf}) &= \mathbb{P}\{\Gamma \leq \gamma_{\rm usf} \bigm | \mbox{UE is a USF-MUE}\} \nonumber\\
&= \int_{0}^{\gamma_{\rm usf}} f_{\Gamma}(\gamma) \,{\rm d}\gamma = \int_{0}^{\gamma_{\rm usf}} \int_{0}^{\min\left(\frac{\gamma}{\tau}, \frac{1}{\gamma}\right)} f_{\Gamma,\Gamma'}(\gamma,\gamma') \,{\rm d}\gamma' \,{\rm d}\gamma, \label{eq:cdf_usfsir}
\end{align}
for $0 \leq \gamma_{\rm usf} \leq \rho$ and, the CDF of throughput of the USF-MUEs can be derived as a function of $F_\Gamma(\gamma_{\rm usf})$ in \eqref{eq:cdf_usfsir} as,
\begin{align}
{\rm F}_{C_{\rm usf}}(c_{\rm usf}) &= \mathbb{P}\{C_{\rm USF} \leq c_{\rm usf} \bigm | \mbox{UE is a USF-MUE}\} \nonumber\\
&= \mathbb{P}\{\log_2(1+\Gamma_{\rm usf}) \leq c_{\rm usf} \bigm | \mbox{UE is a USF-MUE}\},\nonumber\\
&= \mathbb{P}\{\Gamma_{\rm usf} \leq (2^{c_{\rm usf}}-1) \bigm | \mbox{UE is a USF-MUE}\} \nonumber\\
&= {\rm F}_{\Gamma}(2^{c_{\rm usf}}-1),
\label{eq:Cap_usf_cdf}
\end{align}
for $0 \leq c_{\rm usf} \leq \log_2(1+\rho)$. By using the CDF plots, the 5th percentile throughput of USF-MUEs can easily be found as the value at which the CDF is equal to 0.05. Similarly, the 5th percentile throughput of other three UE categories can also be found.

\section{Numerical and Simulation Results}
The average SE and 5th percentile throughput expressions derived in the earlier sections are validated using a Monte Carlo simulation model built in Matlab. Validation of the PPP capacity results for a HetNet scenario with range expansion and reduced power subframes is a non-trivial task. In this section, details of the simulation approach used for validating the PPP analyses are explicitly documented to enable reproducibility. Matlab codes for the simulation model and the theoretical analysis can be downloaded from \cite{MpactWebsite}.
\begin{figure}[h!]
\vspace{-3mm}
\center
\includegraphics[width=6in]{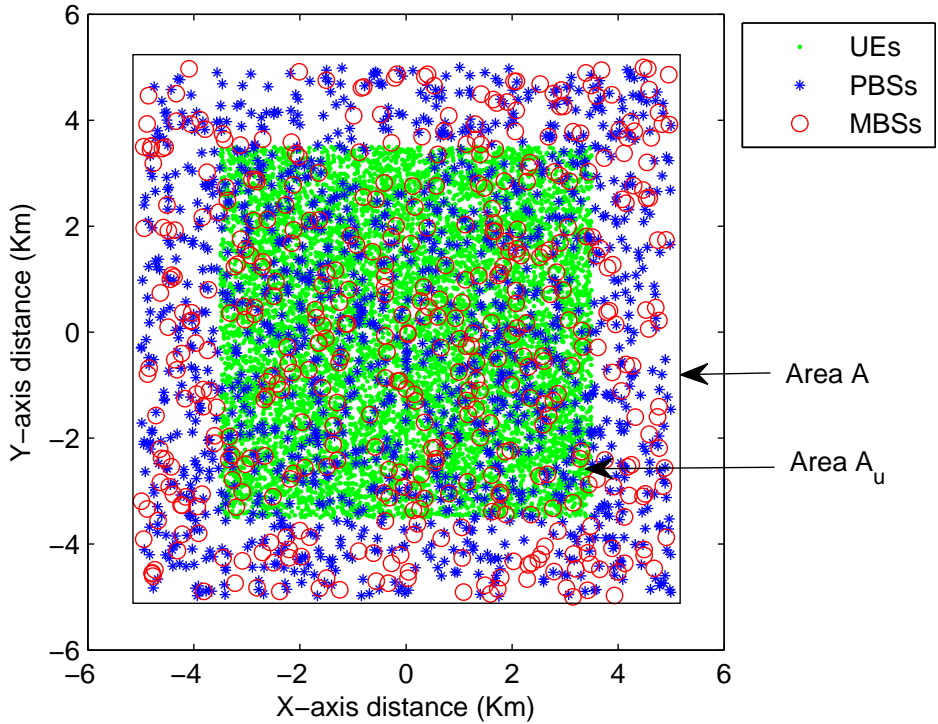}
\vspace{-12mm}
\caption{Simulation layout.}
\label{fig:SimLayout}
\vspace{-5mm}
\end{figure}
\subsection{Simulation Methodology for verifying PPP Model}
Algorithm used in simulation to find the aggregate and per-user SEs is described below.
\begin{enumerate}
\item The X- and Y-coordinates of MBSs, PBSs and UEs are generated using uniformly distributed random variables. The number of MBS and PBS location marks are $\lambda A$ and $\lambda' A$ respectively, where, $A$ is the assumed geographical area that is square in shape as illustrated in Figure~\ref{fig:SimLayout}.
\item The UE locations are constrained within a smaller area $A_{\rm u}$ which is aligned at center of the main simulation area $A$ to avoid the UEs to be located at the edges. In the PPP analysis, the area is assumed to be infinite. But in simulation, this scenario is approximated by making $A$ sufficiently larger than $A_{\rm u}$. The number of UEs is $\lambda_{\rm u}A_{\rm u}$.
\item The MOI (closest MBS) and POI (closest PBS) for each UE is identified. The minimum distance constraints are applied by discarding the UEs that are closer than $d_{\rm min} (d'_{\rm min})$ from their respective MOIs (POIs).
\item The SIRs $\Gamma$, $\Gamma'$, $\Gamma_{\rm CSF}$, $\Gamma'_{\rm CSF}$ are calculated for each UE using \eqref{sirmacusf}-\eqref{sirpiccsf}.
\item The UEs are classified as USF-MUEs, CSF-MUEs, USF-PUEs and CSF-PUEs using the conditions in \eqref{usf_mue}-\eqref{csf_pue}.
\item The MUEs (PUEs) which share the same MOI (POI) are grouped together to form the macro- and pico-cells.
\item The SEs of all the UEs are calculated. In a cell, SE of a USF-MUE $i$ is calculated using \\$\beta\log_2(1+\Gamma_i)/\left(\mbox{No. of USF-MUEs in the cell}\right)$. The SEs of other UE types are calculated using similar formulations.
\item The aggregate capacity of each UE type is calculated in all the cells.
\item Mean aggregate capacity and mean number of UEs of each type are calculated by averaging over all the cells.
\item The per-user SE of each UE type are calculated by (mean aggregate capacity)/(mean number of UEs).
\end{enumerate}

\begin{table}[h!]
\caption{Parameter settings.}
\label{tab:ParamSet}
\center
\begin{tabular}{|l|l|}
\hline
$P, P'$ & 46 dBm, 30 dBm\\ \hline
$K, K'$ & -11 dBm\\ \hline
$d_{\rm min}, d'_{\rm min}$ & 35 m, 10 m \\ \hline
$\lambda, \lambda', \lambda_{\rm u}$ (marks/Km$^2$) & 4.6, $3\lambda$, 200 \\ \hline
Fading model, Path-loss exponent ($\delta$) & Rayleigh, 4\\ \hline
$\beta, \rho, \rho'$ & 0.5, 4 dB, 0dB\\ \hline
\end{tabular}
\end{table}

\subsection{Per-user SEs with PPPs and Monte Carlo Simulations}
The system parameter settings are shown in Table~\ref{tab:ParamSet}. The per-user SE results obtained using analytic expressions of \eqref{eq:perUSF_MUE_SE}-\eqref{eq:perCSF_PUE_SE} are shown in Figure~\ref{fig:Per_user_SE}(a) and Figure~\ref{fig:Per_user_SE}(b) for macrocell and picocell, respectively. The analytic plots agree well with the simulation plots, and provide the following insights:
\begin{figure}[h!]
\center
\includegraphics[width=5in]{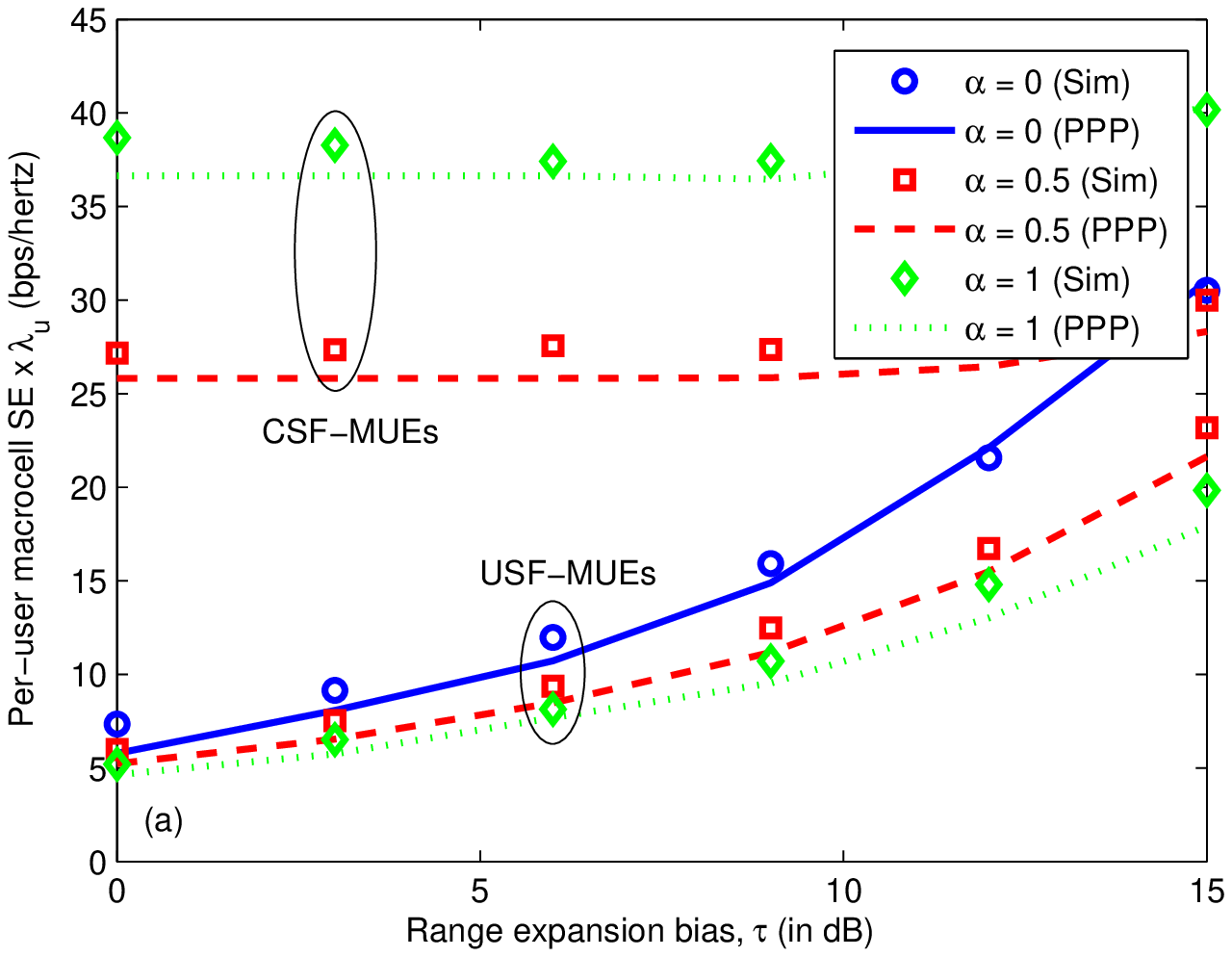}
\includegraphics[width=5in]{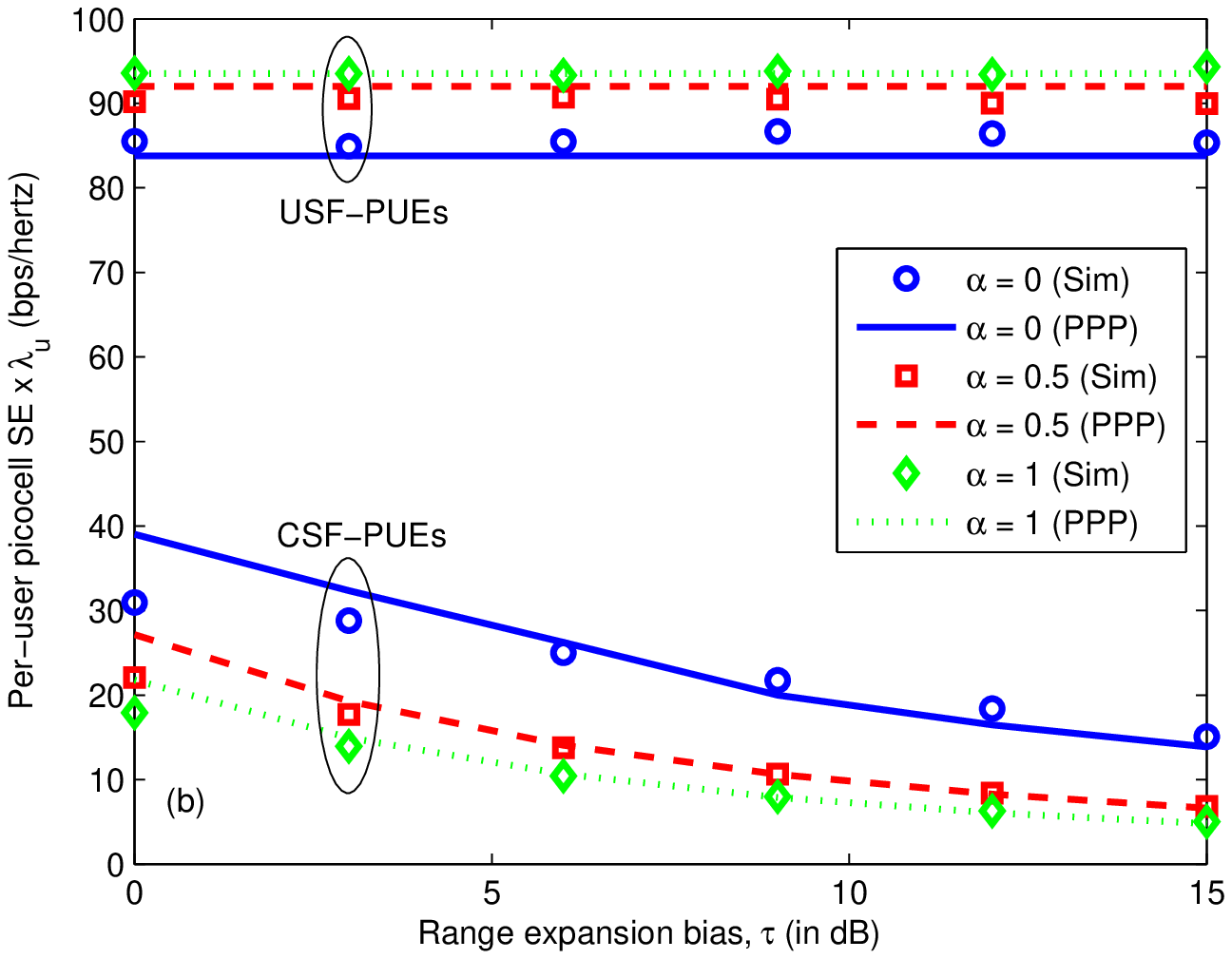}
\vspace{-6mm}
\caption{Per-user SE in (a)~macrocell; (b)~picocell. For the case with $\beta$ = 0.5, $\rho$ = 4~dB and $\rho'$ = 0~dB.}
\label{fig:Per_user_SE}
\end{figure}

\subsubsection{USF- and CSF-MUEs}
Referring to Figure~\ref{fig:SIR}, USF-MUEs form the outer part and CSF-MUEs form the inner part of the macrocell. As the REB increases, some of the USF-MUEs at the macro-pico boundary which have worse SIRs are offloaded to the picocell. Consequently, the mean number of USF-MUEs decreases and their per-user SE increases as shown in Figure~\ref{fig:Per_user_SE}(a).

The mean number of CSF-MUEs are not affected by $\tau$ as long as $\sqrt{\tau} \leq \rho$. Considering Figure~\ref{fig:Capacity_int_regions}, it can be noted that if $\sqrt{\tau}=\rho$, the line $\gamma=\tau\gamma'$ intersects the boundary of region R1. Hence, if $\tau$ is increased further such that $\sqrt{\tau} > \rho$, the area of R1 decreases and thereby decreases the mean number of CSF-MUEs. Therefore, the per-user SE of CSF MUEs remains constant as long as $\sqrt{\tau} \leq \rho$, and increases if $\tau$ crosses this limit as shown in Figure~\ref{fig:Per_user_SE}(a).

On the other hand, as the $\alpha$ increases, the transmit power of all the interfering MBSs increases during CSFs, hence it increases the interference power $Z$ at all the UEs. This causes the SIRs of USF-MUEs ($\Gamma$), USF-PUEs ($\Gamma'$) and CSF-PUEs ($\Gamma'_{\rm csf}$) to decrease, which can be noted in \eqref{sirmacusf}, \eqref{sirpicusf}, and \eqref{sirpiccsf}, respectively. However, the SIRs of CSF-MUEs ($\Gamma_{\rm csf}$) would increase (despite of increased interference) because of the increase in received signal power (due to higher $\alpha$) which can be noted in \eqref{sirmaccsf}. Considering \eqref{usf_mue} and \eqref{csf_mue}, since $\rho$ is a constant, the degradation in $\Gamma$ causes the number of USF-MUEs to increase and CSF-MUEs to decrease. Consequently, the per-user SE of USF-MUEs decreases and that of CSF-MUEs increases for increasing $\alpha$, as shown in Figure~\ref{fig:Per_user_SE}(a).

\subsubsection{USF- and CSF-PUEs}
As the REB increases, the mean number of USF-PUEs remains constant if $\rho' > 1/\sqrt{\tau}$ because the area of region R4 in Figure~\ref{fig:Capacity_int_regions} is unaffected by the value of $\tau$. Therefore, the per-user SE of USF-PUEs also remain constant for increasing REB as shown in Figure~\ref{fig:Per_user_SE}(b). With increasing REB, some MUEs are offloaded to the picocell and become CSF-PUEs. But, these UEs are located at cell-edges and have low SIRs. Hence the per-user SE of CSF-PUEs decreases as shown in Figure~\ref{fig:Per_user_SE}(b).

On the other hand, as the $\alpha$ increases, the transmit power of all the interfering MBSs increases during CSFs causing $\Gamma$, $\Gamma'$ and $\Gamma'_{\rm csf}$ to decrease and $\Gamma_{\rm csf}$ to increase, as explained previously. Considering \eqref{usf_pue} and \eqref{csf_pue}, since $\rho'$ is a constant the degradation in $\Gamma'$ causes the number of USF-PUEs to decrease and CSF-PUEs to increase. Consequently, the per-user SE of USF-PUEs increases and that of CSF-PUEs decreases for increasing $\alpha$, as shown in Figure~\ref{fig:Per_user_SE}(b).

\subsection{Optimization of System Parameters to Achieve Maximum Capacity and Proportional Fairness}
\label{sec:optimization}
The five parameters $\tau,\ \alpha,\ \beta,\ \rho,\mbox{ and } \rho'$ are the key system parameters that are critical to the satisfactory performance of the HetNet system. The goal of these parameter settings is to maximize the aggregate capacity in a cell while providing proportional fairness among the users.

Consider an arbitrary cell which consists of $N$ UEs. Let $C_i$ be the capacity of an arbitrary UE $i \in \{1,\ 2, ...,\ N\}$. The sum of capacities (sum-rate) and the sum of log capacities (log-rate) in a cell are respectively given by,
\begin{align}
C_{\rm sum} = \sum_{i = 1}^{N} C_i,\ \ 
C_{\rm log} = \sum_{i = 1}^{N} \log(C_i) = \log\left( \prod_{i=1}^{N} C_i\right). \label{eq:Clog}
\end{align}
Maximizing the $C_{\rm sum}$ corresponds to maximizing the aggregate capacity in a cell, while maximizing the $C_{\rm log}$ corresponds to proportional fair resource allocation to the users of a cell \cite[App.~A]{1003822}, \cite{6362834}. There can be trade-offs existing between aggregate capacity and fairness in a cell. Maximizing the $C_{\rm sum}$ may reduce the $C_{\rm log}$, and vice versa. In this section, we try to understand these trade-offs by analyzing the characteristics of $C_{\rm log}$ and $C_{\rm sum}$ with respect to the variation of key system parameters.

\begin{figure}[h!]
\center
\includegraphics[width=6in]{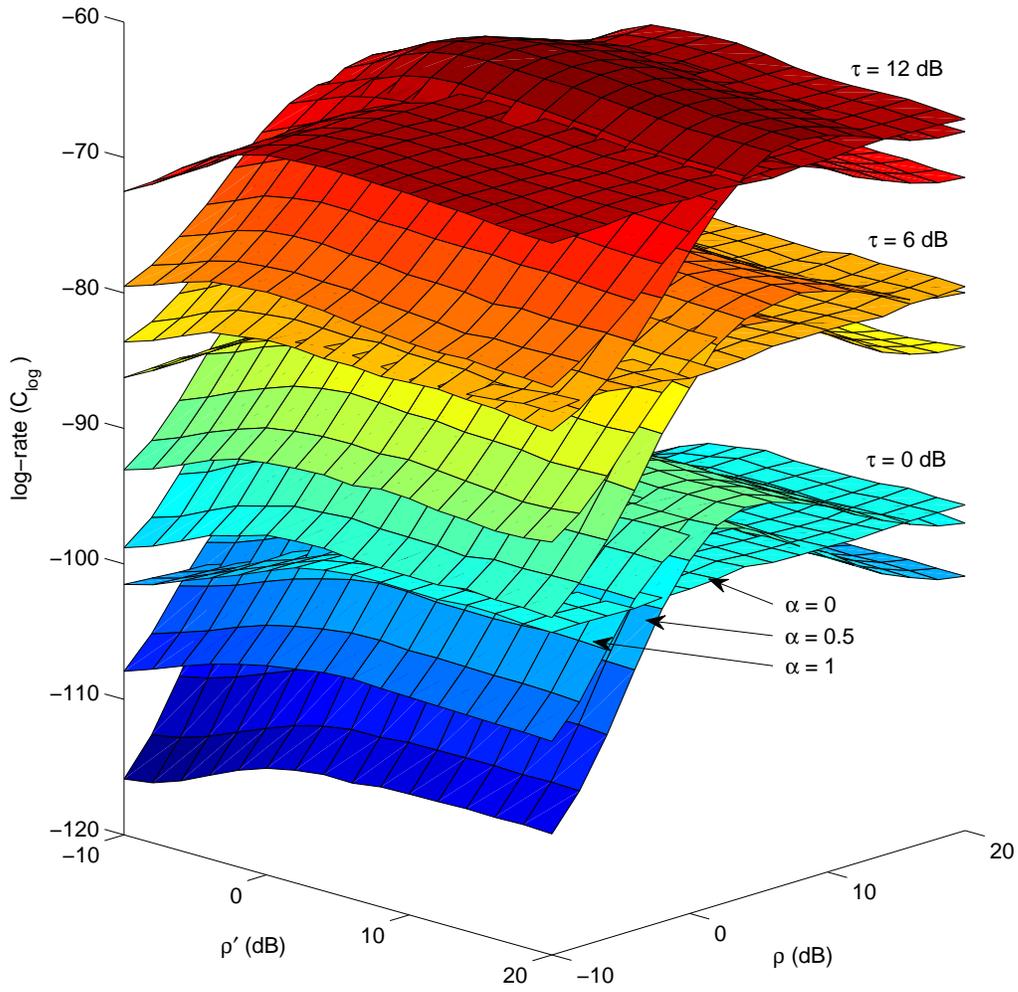}
\caption{Sum of log capacities versus the scheduling thresholds for different $\alpha$ and $\tau$ combinations.}
\label{fig:Clog_all}
\end{figure}
We attempt to achieve the proportional fairness by optimizing the five key system parameters to maximize the $C_{\rm log}$. The variation of $C_{\rm log}$ with respect to $\rho,\ \rho',\ \alpha,\ \tau$ is shown in Figure~\ref{fig:Clog_all}, for $\beta = 0.5$. These plots are obtained through the Monte Carlo simulations and each surface plot is the variation of $C_{\rm log}$ with respect to $\rho$ and $\rho'$ for a fixed value of $\alpha$ and $\tau$. The optimum scheduling thresholds $\rho^*$ and $\rho'^*$ that maximizes the $C_{\rm log}$ are dependent on the values of $\alpha$ and $\tau$.

\begin{figure}[h!]
\center
\includegraphics[width=5in]{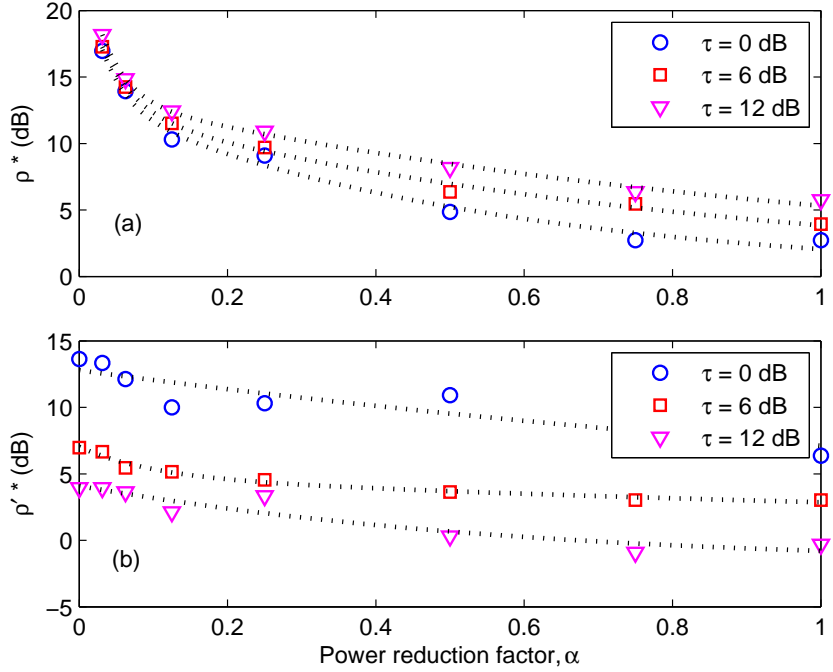}
\vspace{-9mm}
\caption{Optimized scheduling thresholds versus $\alpha$ for different $\tau$ (a) in macrocell; (b) in picocell. With $\lambda = 4.6$~marks/Km$^2$ and $\lambda' = 13.8$~marks/Km$^2$.}
\label{fig:OptimumSchedulingThresholds_separate}
\end{figure}
Figure~\ref{fig:OptimumSchedulingThresholds_separate} shows the plots of $\rho^*$ and $\rho'^*$ as the functions of $\alpha$ and $\tau$. The markers show the simulation results while the dotted lines show the smoother estimation obtained using the curve fitting tool in MATLAB. For small $\alpha$ values, the optimum threshold $\rho^*$ has higher values as shown in Figure~\ref{fig:OptimumSchedulingThresholds_separate}(a), and according to \eqref{csf_mue} this causes very few MUEs that have $\Gamma > \rho^*$ to be scheduled during CSFs. This makes sense because MBS transmit power during CSFs is very low for small $\alpha$ and hence the number of CSF-MUEs which can be covered is also less. On the other hand, for higher $\alpha$ values, MBS transmits with higher power level during CSFs and can cover larger number of CSF-MUEs. Therefore, to improve the fairness proportionally, the optimal $\rho^*$ value decreases with increasing $\alpha$ so that more MUEs are scheduled during CSFs.

In the picocell, with increasing $\alpha$ the CSF-PUEs at the cell edges will experience higher interference from the MBSs. Then, more PUEs should be scheduled during USFs to improve proportional fairness. Likewise, decreasing $\rho'^*$ in Figure~\ref{fig:OptimumSchedulingThresholds_separate}(b) indicates that more PUEs are scheduled during USFs as per \eqref{usf_pue}.

\begin{figure}[h!]
\center
\includegraphics[width=5in]{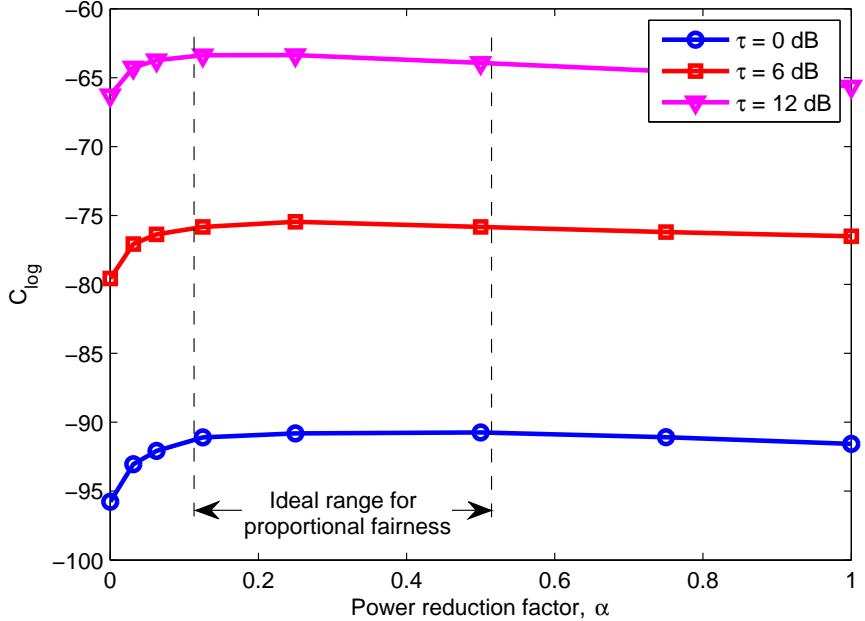}
\vspace{-5mm}
\caption{$C_{\rm log}$ versus $\alpha$ with optimum scheduling thresholds $\rho^*$ and $\rho'^*$. With $\lambda = 4.6$~marks/Km$^2$ and $\lambda' = 13.8$~marks/Km$^2$.}
\label{fig:Clog_max_separate}
\end{figure}
The $C_{\rm log}$ with optimum scheduling thresholds $\rho^*$ and $\rho'^*$ is plotted in Figure~\ref{fig:Clog_max_separate}. Higher the $C_{\rm log}$, better is the proportional fairness. It is important to note that the range expansion bias, $\tau$, has a significant effect on proportional fairness. The $C_{\rm log}$ increases from $-40$ to $-28$ when $\tau$ is increased from 0 db to 12 dB.

Compared to $\tau$, $\alpha$ has a smaller effect on the proportional fairness. When $\alpha$ is set to zero which corresponds to the eICIC, $C_{\rm log}$ is at its minimum. It shows that eICIC provides minimum proportional fairness. Figure~\ref{fig:Clog_max_separate} moreover shows that setting $\alpha = 1$ which corresponds to no eICIC, also does not provide maximum $C_{\rm log}$. An $\alpha$ setting between 0.125 and 0.5 maximizes the $C_{\rm log}$ and hence the proportional fairness.

\begin{figure}[h!]
\center
\includegraphics[width=5in]{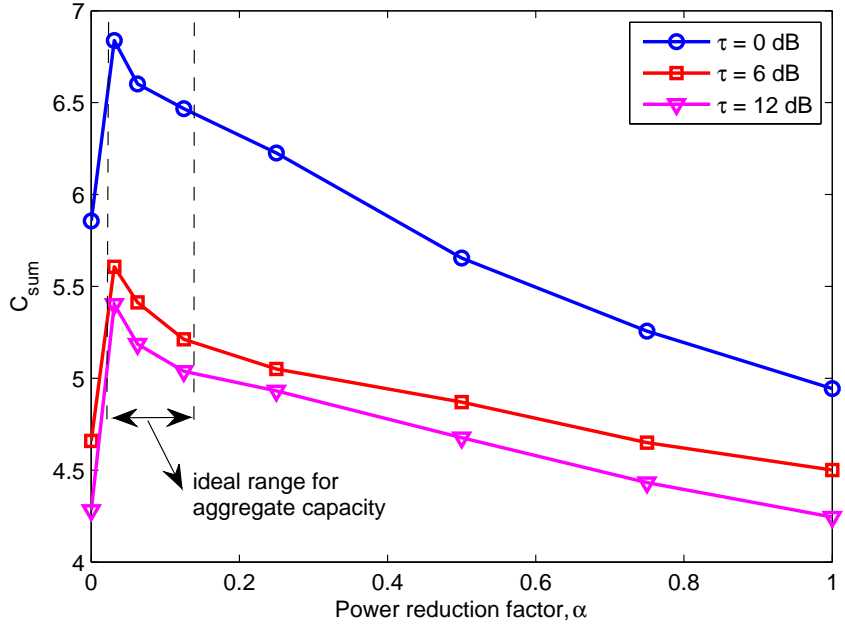}
\vspace{-5mm}
\caption{$C_{\rm sum}$ versus $\alpha$ with optimum scheduling thresholds $\rho^*$  and $\rho'^*$. With $\lambda = 4.6$~marks/Km$^2$ and $\lambda' = 13.8$~marks/Km$^2$.}
\label{fig:Csum_opt}
\end{figure}
The characteristics of $C_{\rm sum}$ with optimum scheduling thresholds is shown in Figure~\ref{fig:Csum_opt}. As the $\tau$ increases, $C_{\rm sum}$ decreases, which is the opposite effect when compared to the $C_{\rm log}$ in Figure~\ref{fig:Clog_max_separate}. This shows the trade-off between the aggregate capacity and the proportional fairness. Increasing the $\tau$ would increase the proportional fairness but decrease the aggregate capacity, and vice versa.

Comparing Figures~\ref{fig:Clog_max_separate} and \ref{fig:Csum_opt} also explains the trade-off associated with setting $\alpha$. A very small value, $0 < \alpha < 0.125$, provides larger $C_{\rm sum}$ but smaller $C_{\rm log}$, which is better from an aggregate capacity point of view. Setting $0.125 \leq \alpha \leq 0.5$ is better from a fairness point of view. Any value of $\alpha > 0.5$ is not recommended since it degrades the aggregate capacity as shown in Figure~\ref{fig:Csum_opt}, decreases the proportional fairness as shown in Figure~\ref{fig:Clog_max_separate}, and consumes higher transmit power by the MBSs. Setting $\alpha = 0$ as in the eICIC case would reduce both $C_{\rm sum}$ and $C_{\rm log}$ drastically.

\begin{figure}[h!]
\center
\includegraphics[width=5in]{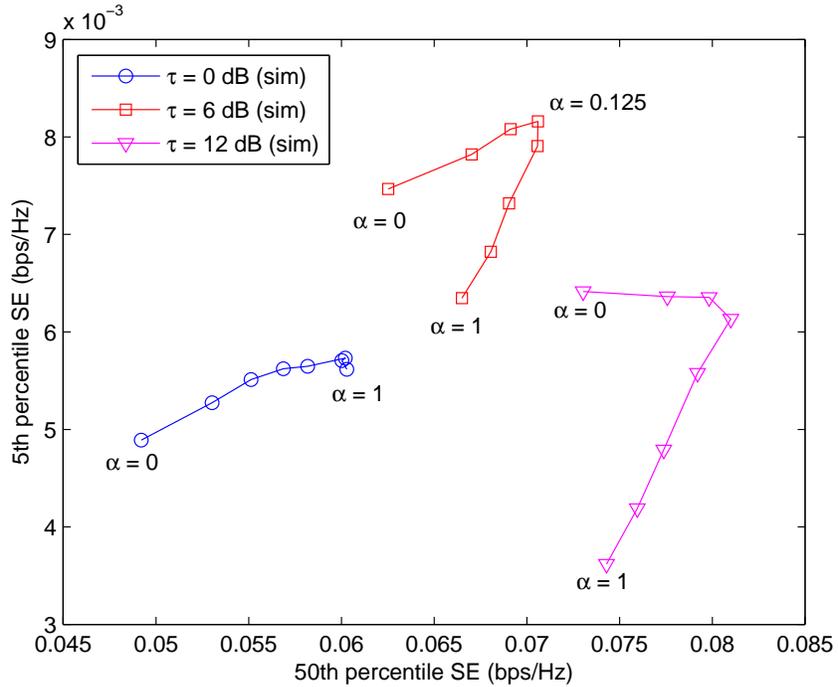}
\vspace{-5mm}
\caption{5th percentile capacity versus 50th percentile capacity. With $\lambda = 4.6$~marks/Km$^2$ and $\lambda' = 13.8$~marks/Km$^2$.}
\label{fig:5th_50th_opt_ro_ro1}
\end{figure}
The implications in Figures~\ref{fig:Clog_max_separate} and \ref{fig:Csum_opt} can be seen from a different perspective by using the 5th percentile SE versus 50th percentile SE graph shown in Figure~\ref{fig:5th_50th_opt_ro_ro1}. It shows that increasing the $\tau$ from 6 dB to 12 dB improves the 50th percentile SE, but degrades the 5th percentile SE. This illustrates the trade-off between the 5th and 50th percentile SEs, which is analogous to the trade-off between aggregate capacity and proportional fairness among the users, as explained in the previous paragraphs. Figure~\ref{fig:5th_50th_opt_ro_ro1} also shows that $\alpha$ has a notable effect on the 5th and 50th percentile SEs.

\subsection{Impact of the Duty Cycle of Uncoordinated Subframes}
\begin{figure}[h!]
\center
\includegraphics[width=5in]{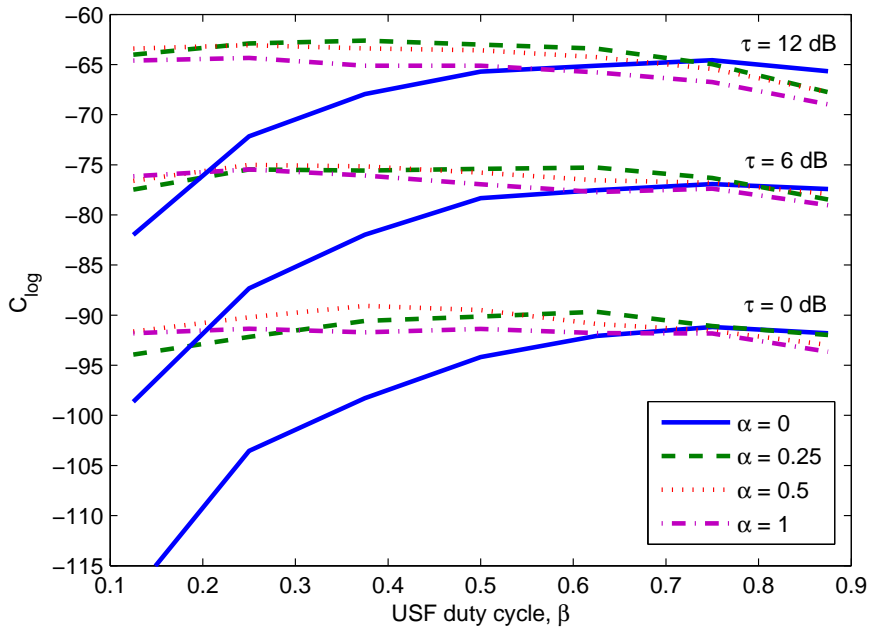}
\vspace{-5mm}
\caption{$C_{\rm log}$ versus $\beta$ with optimum scheduling thresholds $\rho^*$  and $\rho'^*$. With $\lambda = 4.6$~marks/Km$^2$ and $\lambda' = 13.8$~marks/Km$^2$.}
\label{fig:Clog_ro_ro1_opt}
\end{figure}

\begin{figure}[h!]
\center
\includegraphics[width=3.5in]{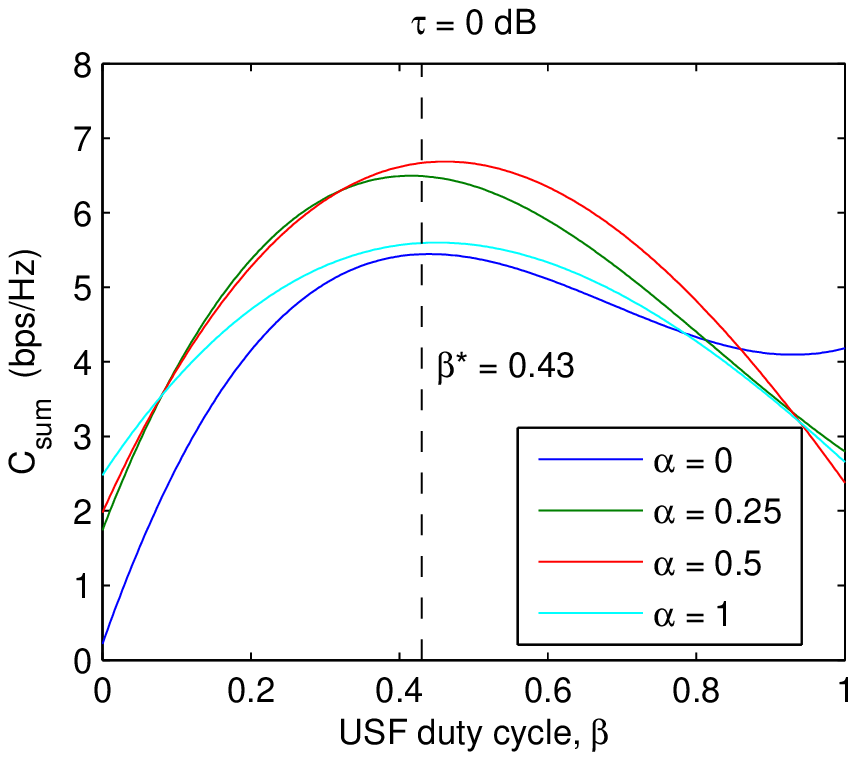}
\includegraphics[width=3.5in]{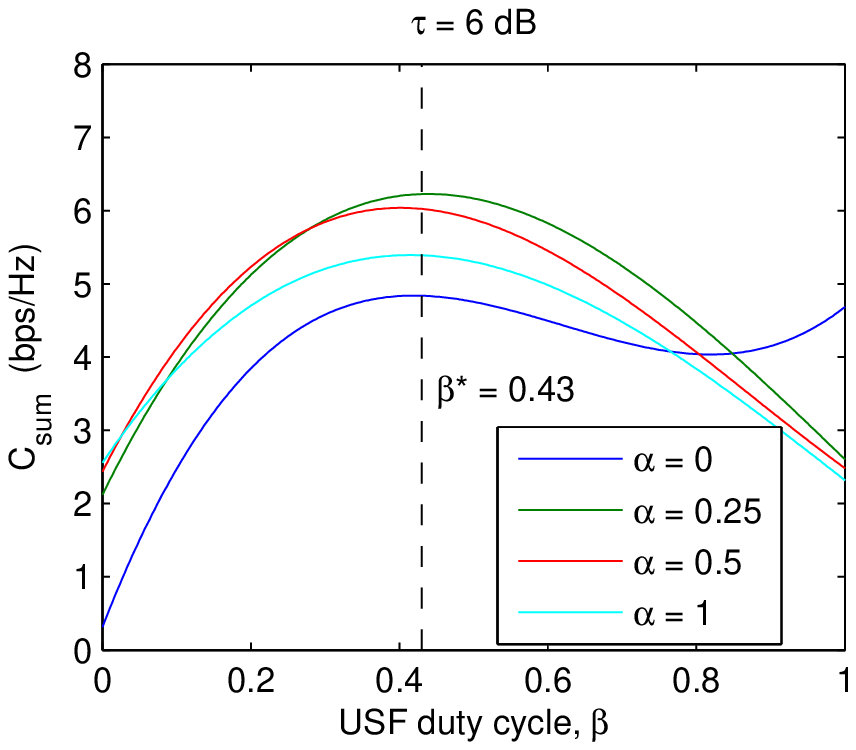}
\includegraphics[width=3.5in]{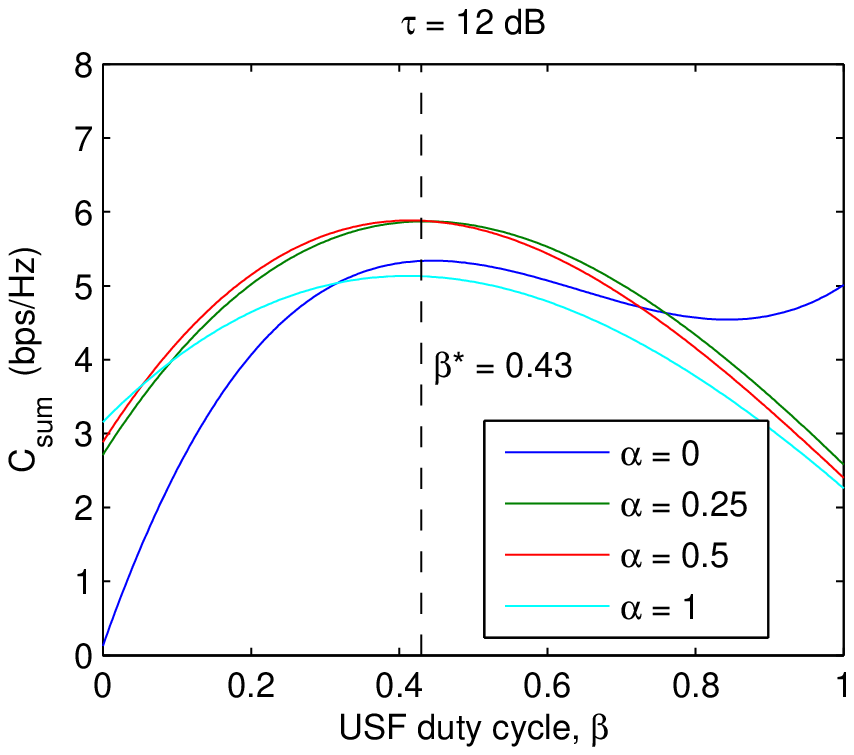}
\vspace{-4mm}
\caption{$C_{\rm sum}$ ve1rsus $\beta$ with optimum scheduling thresholds $\rho^*$  and $\rho'^*$. With $\lambda = 4.6$~marks/Km$^2$ and $\lambda' = 13.8$~marks/Km$^2$.}
\label{fig:Csum_ro_ro1_opt}
\end{figure}

In the results of Figures~\ref{fig:OptimumSchedulingThresholds_separate}--\ref{fig:5th_50th_opt_ro_ro1}, $\beta$ was set to 0.5 and we next show the effect of varying $\beta$ on $C_{\rm log}$ and $C_{\rm sum}$. Introducing $\beta$ into the optimization problem makes it difficult to visualize the results due to the addition of one more dimension. Therefore, we use the optimized scheduling thresholds, $\rho^*$ and $\rho'^*$, and analyze $C_{\rm log}$ and $C_{\rm sum}$ as the functions $\beta$, $\alpha$ and $\tau$. Figures~\ref{fig:Clog_ro_ro1_opt} and \ref{fig:Csum_ro_ro1_opt} show the $C_{\rm log}$ versus $\beta$ and the $C_{\rm sum}$ versus $\beta$, respectively for different values of $\alpha$ and $\tau$. The variation of $C_{\rm log}$ with respect to $\beta$ is not significant, except for $\alpha = 0$. Whereas, the variation of $C_{\rm sum}$ with respect to $\beta$ is significant.

When $\alpha = 0$, the $C_{\rm log}$ value decreases rapidly for $\beta < 0.5$. Nevertheless, $\alpha = 0$ is shown to have poor performance in the previous paragraphs and hence it is not recommended. For other values of $\alpha$, variation in $\beta$ does not affect the $C_{\rm log}$ significantly, which shows that by using a fixed value of $\beta$, proportional fairness can be achieved by optimizing (to maximize $C_{\rm log}$) the scheduling thresholds. Figure~\ref{fig:Csum_ro_ro1_opt} shows that fixing $\beta$ approximately to 0.43 maximizes the $C_{\rm sum}$ irrespective of $\alpha$ and $\tau$, provided the scheduling thresholds are optimized to maximize $C_{\rm log}$.

In \cite{Jeff_2013_arxiv}, the boundary of CSF-PUEs that form the inner region of picocell (excluding the range expansion region) is fixed due to the fixed transmit power of PBS. The \emph{association bias} and \emph{resource partitioning fraction} parameters are used as the variables to be optimized. It is analogous for us to have a fixed $\rho'$ and optimize $\beta$ and $\tau$. But in contrast, we fix the $\beta$ for simplicity and optimize the other four parameters, since coordinating $\beta$ among the cells through the X2 interface is complex and adds to communication overhead in the backhaul.

\subsection{5th Percentile Throughput}
\begin{figure}[h!]
\center
\includegraphics[width=5in]{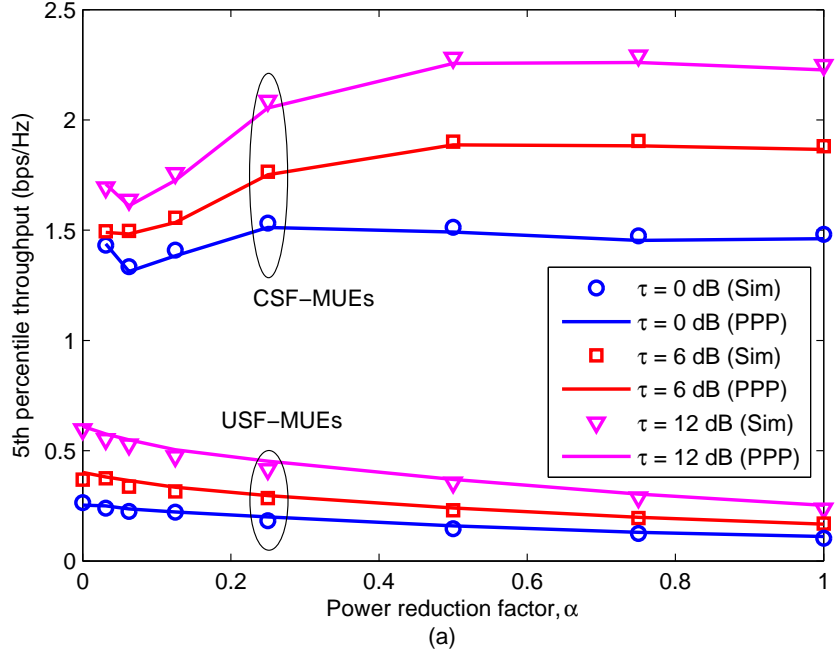}
\includegraphics[width=5in]{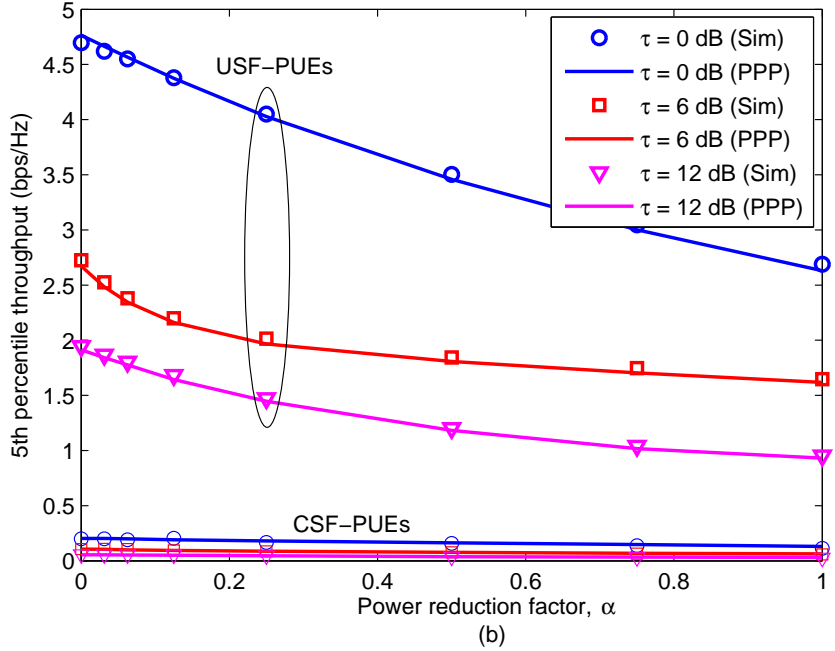}
\vspace{-5mm}
\caption{5th percentile throughput (a) in macrocell; (b) in picocell. With $\lambda = 4.6$~marks/Km$^2$ and $\lambda' = 13.8$~marks/Km$^2$.}
\label{fig:5_perctile_vs_alpha}
\end{figure}

Using the expressions derived in Section~\ref{sec:5th_perc_thrpt}, the 5th percentile throughput versus $\alpha$ for different $\tau$ is shown in Figure~\ref{fig:5_perctile_vs_alpha}(a) for MUEs, and in Figure~\ref{fig:5_perctile_vs_alpha}(b) for PUEs. As the $\alpha$ increases, MBSs transmit at higher power level during CSFs and the UEs of all types experience a higher interference power. However, the received signal power at CSF-MUEs increases with $\alpha$ and results in improved 5th percentile throughput as shown in Figure~\ref{fig:5_perctile_vs_alpha}(a). But, the SIRs of USF-MUEs and USF/CSF-PUEs degrade due to higher interference and therefore their 5th percentile throughput decreases with increase in $\alpha$ as shown in Figures~\ref{fig:5_perctile_vs_alpha}(a) and \ref{fig:5_perctile_vs_alpha}(b).

Increasing the REB, $\tau$, causes the USF-MUEs with poor SIR, located at the edge of macrocell, to be offloaded to the picocell and thereby increasing the 5th percentile throughput of USF-MUEs as shown in Figure~\ref{fig:5_perctile_vs_alpha}(a). The offloaded UEs in picocell are scheduled during CSFs and due to their poor SIR the 5th percentile throughput of CSF-PUEs decreases as shown in Figure~\ref{fig:5_perctile_vs_alpha}(b).

\subsection{Comparison with Real BS Deployment}
We obtained the data of real BS locations in United Kingdom from an organization \cite{SiteFinder} where the mobile network operators have voluntarily provided the information of location and operating characteristics of individual BSs. The data set in \cite{SiteFinder} was last updated in May 2012, and it provides exact locations of the BSs. Also, the BSs of different operators can be distinguished.
\begin{figure}[h!]
\center
\includegraphics[width=5in]{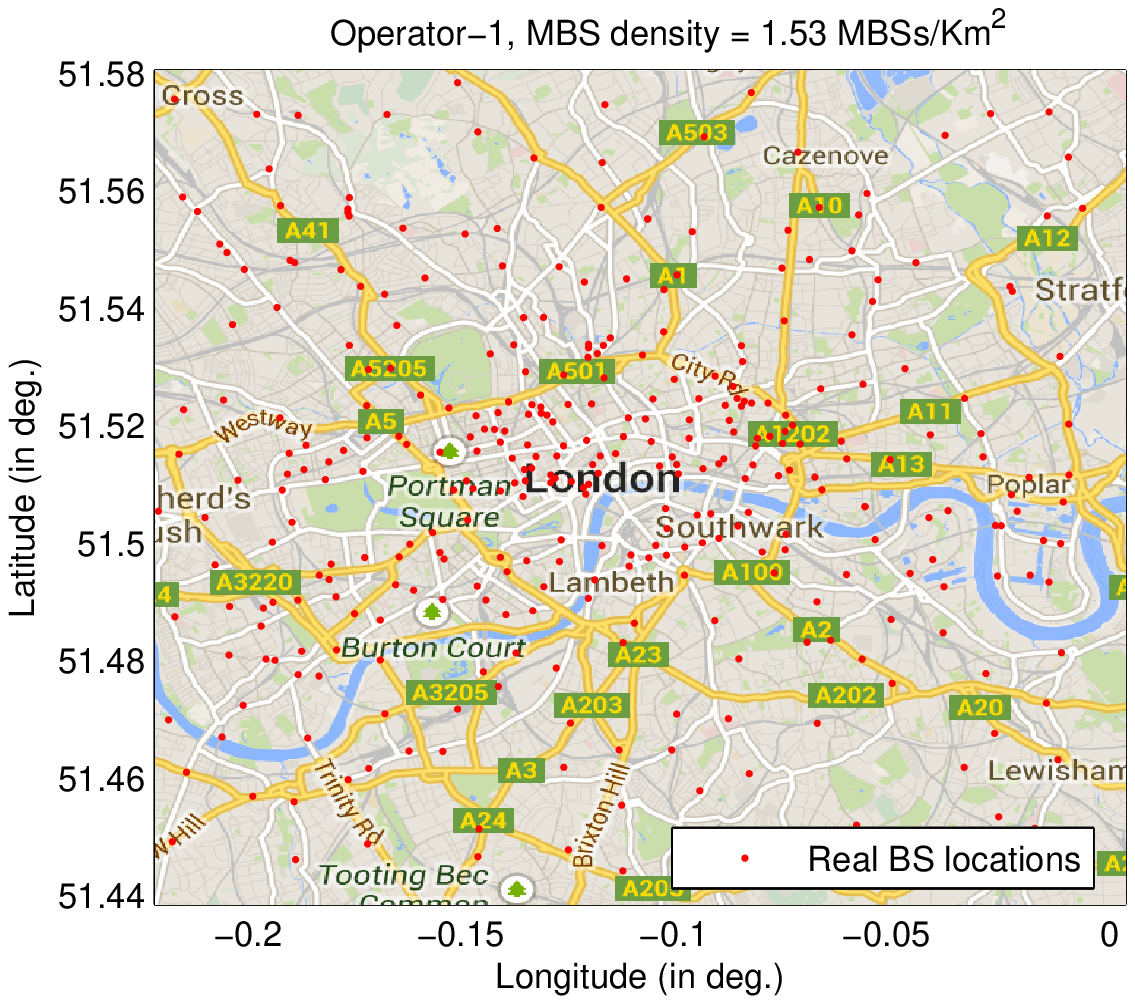}
\includegraphics[width=5in]{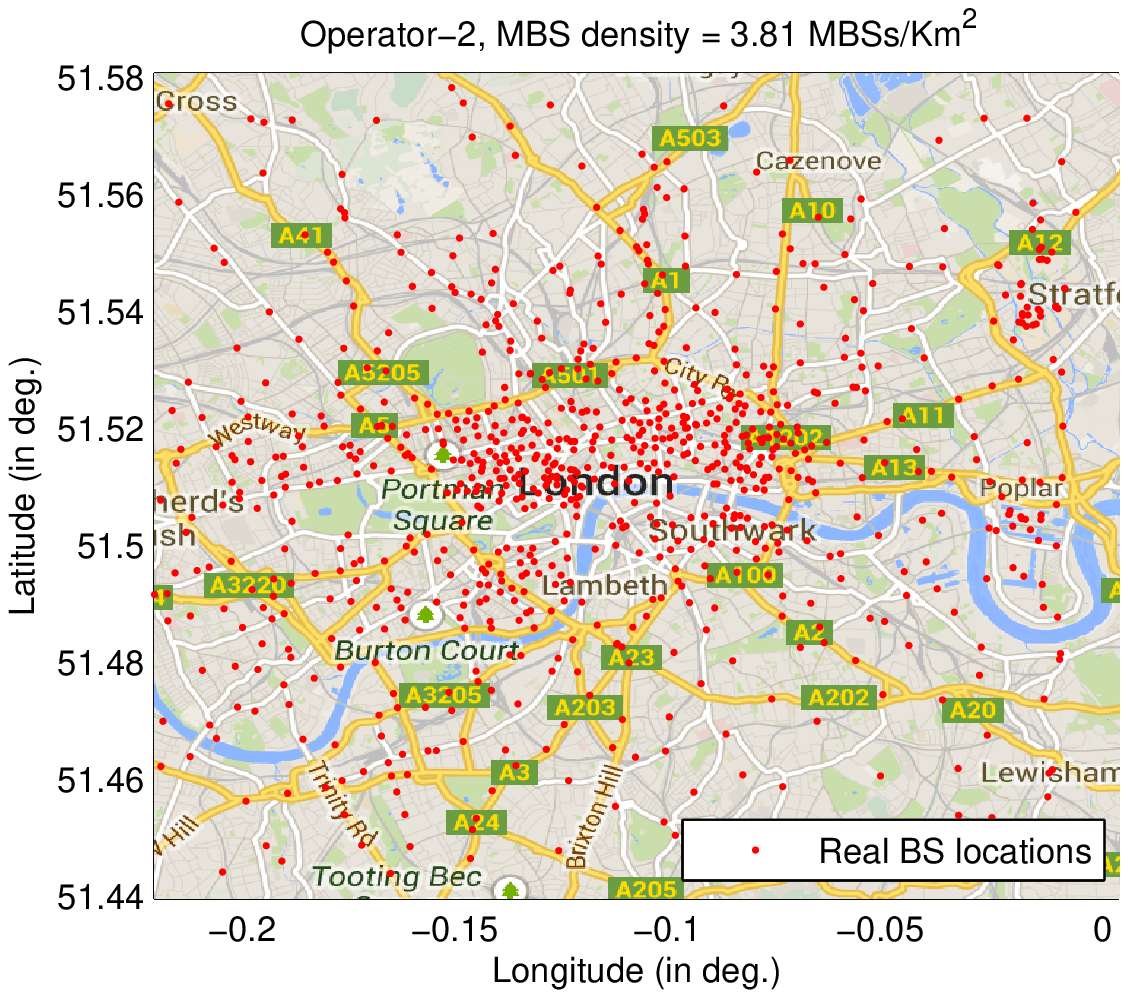}
\vspace{-5mm}
\caption{Real base station locations of two different operators in a $15\times15$ km$^2$ area of London city.}
\label{fig:Real_BS_Layout}
\end{figure}

In this section, we compare the 5th percentile SE results from the PPP model with that of the real BS deployment and hexagonal grid model. The real MBS locations of two different operators in a $15\times15$ km$^2$ area of London city were obtained from \cite{SiteFinder} as shown in Figure~\ref{fig:Real_BS_Layout}. In this area, the average BS densities of the two operators were found to be 1.53~MBSs/km$^2$ and 2.04~MBSs/km$^2$. To have a fair comparison, the MBS locations for hexagonal grid and PPP models were also generated with the same densities. The PBS locations were generated randomly using another PPP model. The parameters $\tau=6$~dB, $\alpha=0.5$, $\beta=0.5$, $\rho=4$~dB, $\rho'=12$~dB, and $P_{\rm tx}=46$~dBm were fixed while the PBS density $\lambda'$ was varied to analyze its effect on the 5th percentile SE.

\begin{figure}[h!]
\center
\includegraphics[width=5in]{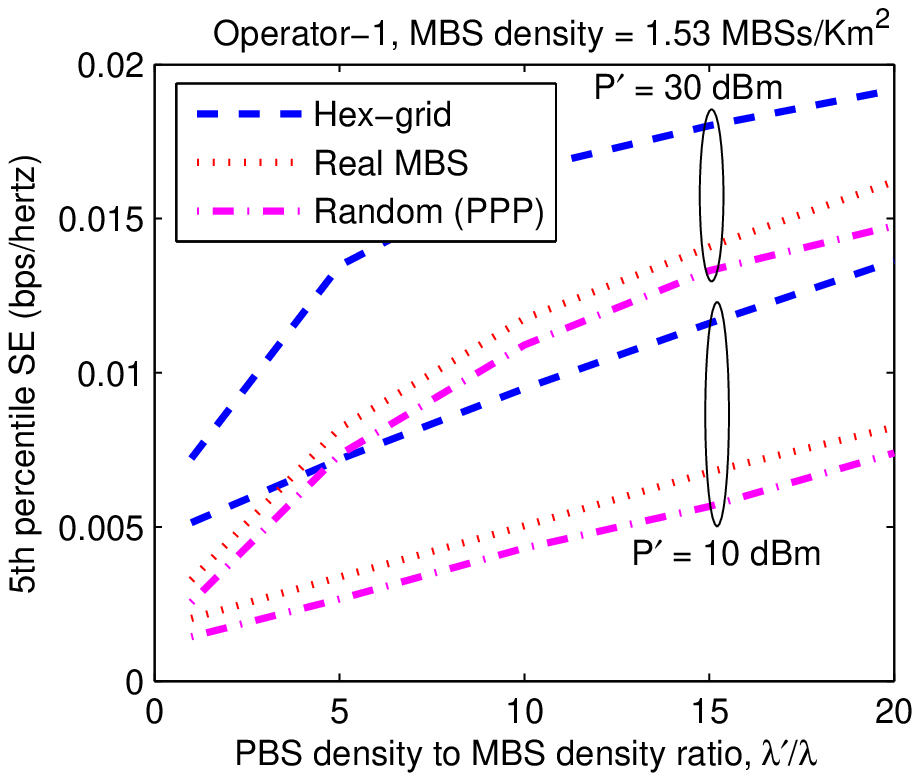}
\includegraphics[width=5in]{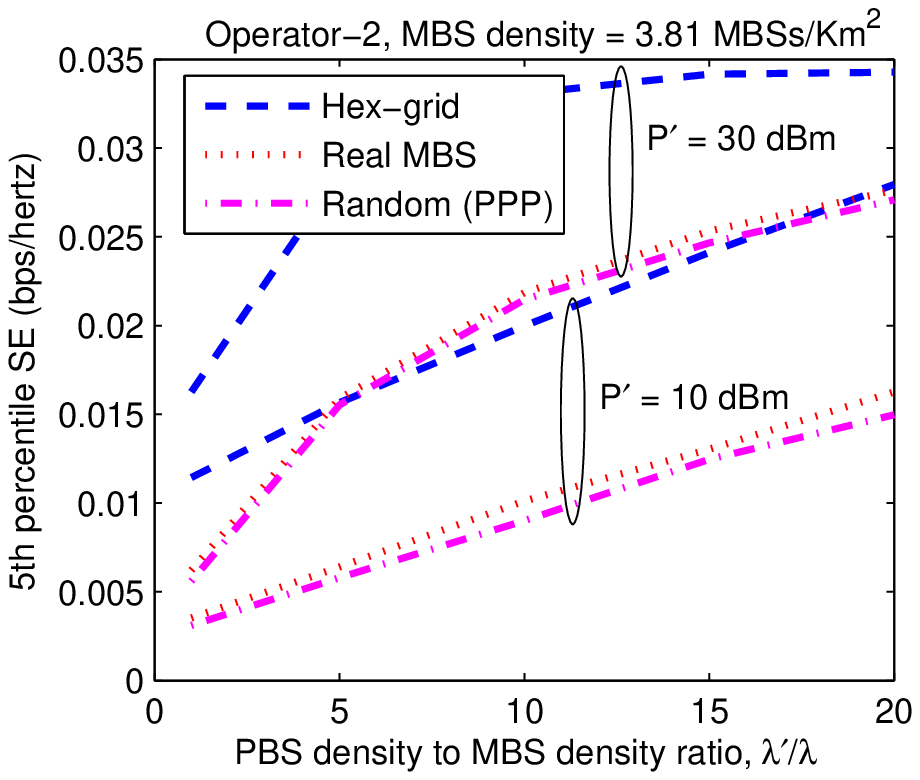}
\vspace{-4mm}
\caption{5th percentile SE versus PBS density.}
\label{fig:RealBS_5pSE_vs_PBS_density}
\end{figure}
The plots of 5th percentile SE versus PBS density are shown in Figure~\ref{fig:RealBS_5pSE_vs_PBS_density} for the two operators. The 5th percentile SE of operator-2 is better than that of operator-1 since the former has higher MBS density. As expected, the 5th percentile SE improves with the increase in PBS density. It can also be observed that increasing the PBS transmit power $P'$ from 10~dBm to 30~dBm will result into almost twice the 5th percentile SE. Since hexagonal grid model is an ideal case, it has the best 5th percentile SE and forms an upper bound. The PPP model has the worse 5th percentile SE and forms a lower bound. The real MBS deployment is usually planned and hence it is not completely random in nature. On the other hand, it is also not equivalent to the idealized hexagonal grid model due to the practical constraints involved during the deployment. Hence, the 5th percentile SE of real MBS deployment lies in between the two bounds of hexagonal grid and random deployments.

\section{Conclusion}
In this paper, spectral efficiency and 5th percentile throughput expressions are derived for HetNets with reduced power subframes and range expansion. These expressions are validated using the Monte Carlo simulations. Joint optimization of the key system parameters, such as range expansion bias, power reduction factor, scheduling thresholds, and duty cycle of reduced power subframes, is performed to achieve maximum aggregate capacity and proportional fairness among users. Our analysis shows that under optimum parameter settings, the HetNet with reduced power subframes yields better performance than that with almost blank subframes (eICIC) in terms of both aggregate capacity and proportional fairness. However, transmitting the reduced power subframes with greater than half the maximum power proved to be inefficient because it degrades both the aggregate capacity and the proportional fairness. Increasing the range expansion bias improves the proportional fairness but degrades the aggregate capacity. In case of eICIC, the duty cycle of almost blank subframes has a significant effect on the fairness, but with reduced power subframes and optimized scheduling thresholds, duty cycle has a limited effect on fairness. Hence, fixing the duty cycle and optimizing the scheduling thresholds is preferable since it avoids the overhead of coordinating the duty cycle among the cells through the X2 interface. We also compared the 5th percentile SE results from PPP model with that of real BS deployment and hexagonal grid model. We observed that the hex grid model forms the upper bound while the PPP model forms the lower bound. Increasing the PBS density or the PBS transmit power would improve the 5th percentile SE.

\newpage
\begin{appendices}
\renewcommand{\thesection}{\arabic{section}}

\section{Derivation of JCCDF Expression} \label{App:JCCDFDerive}
This part of the appendix derives closed form equation for the JCCDF in \eqref{JCCDF_int}. Let us start by rewriting the JCCDF expression,
\begin{align}
\mathbb{P}\{\Gamma > \gamma,\Gamma' > \gamma'| R=r, R'=r'\} = \mathbb{E}_Z \left[ \int_{y1}^{+\infty}f_{\rm Y}(y) \int_{\gamma(y+Z)}^{y/\gamma'-Z} f_{\rm X}(x) \,{\rm d}x\,{\rm d}y \right]
\label{JCCDF1}
\end{align}
where, 
\begin{align}
f_{\rm X}(x)&=\lambda_{\rm x}\exp({-\lambda_{\rm x}x}) \hspace{0.5cm} \mbox{and} \hspace{0.5cm} f_{\rm Y}(y)=\lambda_{\rm y}\exp({-\lambda_{\rm y}y}); \\
\lambda_{\rm x}&=\frac{r^\delta}{P} \hspace{2.16cm} \mbox{and} \hspace{1cm} \lambda_{\rm y}=\frac{(r')^\delta}{P'}. \label{eq:mean_exp}
\end{align}
The inner integral in~\eqref{JCCDF1} can be derived as,
\begin{align}
\int_{\gamma(y+Z)}^{y/\gamma'-Z} f_{\rm X}(x) \,{\rm d}x = \exp\left[-\lambda_{\rm x}\gamma(y+Z)\right]-\exp\left[-\lambda_{\rm x}\left(\frac{y}{\gamma'}-Z\right)\right].
\end{align}
Then, the outer integral in~\eqref{JCCDF1} can be derived as,
\begin{align}
\int_{y1}^{+\infty}\!\!\!\!\!\!\!\!\!f_{\rm Y}(y) \int_{\gamma(y+Z)}^{y/\gamma'-Z} \!\!\!\!\!\!\!\!\!f_{\rm X}(x) \,{\rm d}x\,{\rm d}y =& \lambda_{\rm y}\int_{y1}^{+\infty}\!\!\!\!\!\!\!\!\!\!\exp[-\lambda_{\rm y}y-\lambda_{\rm x}\gamma(y+Z)]\,{\rm d}y - \lambda_{\rm y}\int_{y1}^{+\infty}\!\!\!\!\!\!\!\!\!\exp\left[-\lambda_{\rm y}y-\lambda_{\rm x}\left(\frac{y}{\gamma'}-Z\right)\right]\,{\rm d}y \label{int}.
\end{align}
The first term in right hand side (RHS) of \eqref{int} can be evaluated as,
\begin{align}
\lambda_{\rm y}\int_{y1}^{+\infty}\exp[-\lambda_{\rm y}y-\lambda_{\rm x}\gamma(y+Z)]\,{\rm d}y = \frac{1}{1+\frac{\gamma\lambda_{\rm x}}{\lambda_{\rm y}}}\exp\left[\frac{-\lambda_{\rm x}\gamma Z(1+\gamma') - \lambda_{\rm y}\gamma'Z(1+\gamma)}{1-\gamma\gamma'}\right]. \label{first_term}
\end{align}
The second term in RHS of \eqref{int} can be evaluated as,
\begin{align}
\lambda_{\rm y}\int_{y1}^{+\infty}\exp\left[-\lambda_{\rm y}y-\lambda_{\rm x}\left(\frac{y}{\gamma'}-Z\right)\right]\,{\rm d}y = \frac{1}{1+\frac{\lambda_{\rm x}}{\gamma'\lambda_{\rm y}}}\exp\left[\frac{-\lambda_{\rm x}\gamma Z(1+\gamma') - \lambda_{\rm y}\gamma'Z(1+\gamma)}{1-\gamma\gamma'}\right]. \label{second_term}
\end{align}
By substituting \eqref{first_term} and \eqref{second_term} in the first and second terms of \eqref{int} respectively, we get
\begin{align}
\int_{y1}^{+\infty}f_{\rm Y}(y) \int_{\gamma(y+Z)}^{y/\gamma'-Z} f_{\rm X}(x) \,{\rm d}x\,{\rm d}y = \left(\frac{1}{1+\frac{\gamma\lambda_{\rm x}}{\lambda_{\rm y}}}-\frac{1}{1+\frac{\lambda_{\rm x}}{\gamma'\lambda_{\rm y}}}\right) \exp\left[\frac{-\lambda_{\rm x}\gamma Z(1+\gamma') - \lambda_{\rm y}\gamma'Z(1+\gamma)}{1-\gamma\gamma'}\right]. \label{JCCDF_Closed}
\end{align}
Substituting \eqref{JCCDF_Closed} in \eqref{JCCDF1} and using \eqref{eq:mean_exp} we get,
\begin{align}
\lefteqn{\mathbb{P}\{\Gamma > \gamma,\Gamma' > \gamma'| R=r, R'=r'\}}\nonumber\\
&= \left(\frac{1}{1+\gamma\frac{P'}{P}\left(\frac{r}{r'}\right)^\delta}-\frac{1}{1+\gamma'\frac{P}{P'}\left(\frac{r'}{r}\right)^\delta}\right)
\mathbb{E}_Z \left[ \exp\left(-Z\frac{\frac{\gamma (1+\gamma')r^\delta}{P} + \frac{\gamma'(1+\gamma)(r')^\delta}{P'}}{1-\gamma\gamma'}\right) \right]
\end{align}
Using the definition of Laplace transform, $\mathbb{E}_Z \left[\exp(-Zs)\right] = \mathcal{L}_Z(s)$, and further simplification, we get
\begin{align}
\mathbb{P}\{\Gamma > \gamma,\Gamma' > \gamma'| R=r, R'=r'\} = \frac{(1-\gamma\gamma')\mathcal{L}_Z \left(\frac{1}{1-\gamma\gamma'} \left( \frac{\gamma(1+\gamma')r^\delta}{P}+\frac{\gamma'(1+\gamma)(r')^\delta}{P'}\right)\right)}{\left[ 1+\gamma\frac{P'}{P} \left(\frac{r}{r'}\right)^\delta\right] \left[ 1+\gamma'\frac{P}{P'} \left(\frac{r'}{r}\right)^\delta\right]}.
\label{JCCDF3}
\end{align}

\section{Derivation of JPDF Expression} \label{App:JPDFDerive}
Assuming $\delta=4$, the JCCDF expression in \eqref{JCCDF3} can be rewritten as,
\begin{align}
\mathbb{P}\{\Gamma > \gamma,\Gamma' > \gamma'| R=r, R'=r'\} = M_1 M_2,
\label{JCCDF2}
\end{align}
where,
\begin{align}
M_1 =& \frac{1-\gamma\gamma'}{\left[1+\gamma\frac{P'}{P} \left(\frac{r}{r'}\right)^4\right] \left[1+\gamma'\frac{P}{P'} \left(\frac{r'}{r}\right)^4\right]}, \label{M1}\\
M_2 =& \mathcal{L}_Z \left(\frac{1}{1-\gamma\gamma'} \left( \frac{\gamma(1+\gamma')r^4}{P}+\frac{\gamma'(1+\gamma)(r')^4}{P'}\right)\right). \label{M2}
\end{align}
After some tedious but straight forward algebraic steps, it can be shown that
\begin{align}
M_1 =& \frac{1}{1+\gamma\left(\frac{\tilde{a}}{1-\tilde{a}}\right)} + \frac{1}{1+\gamma'\left(\frac{1-\tilde{a}}{\tilde{a}}\right)} -1,
\label{M1final} \\
M_2 =& \exp\Big\{g\Big(\sqrt{\tilde{a}},\beta\tilde{\mu}\Big) + g\Big(\sqrt{\tilde{a}/\alpha},(1-\beta)\tilde{\mu}\sqrt{\alpha}\Big) + g\Big(\sqrt{1-\tilde{a}},1-\tilde{\mu}\Big)\Big\},
\label{M2final}
\end{align}
where, $\tilde{a}=\frac{1}{1+\frac{P}{P'}\left(\frac{r'}{r}\right)^4}$, $\tilde{\mu}=\frac{1}{1+\frac{\lambda'}{\lambda}\sqrt{\frac{P'}{P}}}$. The function $g$ in \eqref{M2final} is defined as
\begin{align}
g(b,\nu) = -\nu cB\left(\frac{\pi}{2}-\tan^{-1}\frac{b}{c}\right),
\end{align}
where,
\begin{align}
B = \frac{\pi r^2}{\sqrt{P\tilde{a}}} \left(\lambda\sqrt{P}+\lambda'\sqrt{P'}\right) \ \ \ \mbox{and}\ \ \ 
c = \sqrt{\frac{\gamma(1+\gamma')\tilde{a} + \gamma'(1+\gamma)(1-\tilde{a})}{1-\gamma\gamma'}}.
\label{eq:temp2}
\end{align}
We can derive the JPDF by differentiating the JCCDF \eqref{JCCDF2} with respect to $\gamma$ and $\gamma'$,
\begin{align}
f_{\Gamma,\Gamma'\bigm|R,R'}(\gamma,\gamma'\bigm|r,r') = \frac{\partial^2}{\partial\gamma \partial\gamma'} M_1 M_2,
\label{JPDF1}
\end{align}
where $M_1$ and $M_2$ are given by \eqref{M1final} and \eqref{M2final}, respectively. By solving \eqref{JPDF1} it can be shown that the conditional JPDF
\begin{align}
f_{\Gamma,\Gamma'\bigm|R,R'}(\gamma,\gamma'\bigm|r,r') =& M_2 h \left(\frac{\partial M_1}{\partial\gamma}\frac{\partial c}{\partial\gamma'} + \frac{\partial M_1}{\partial\gamma'} \frac{\partial c}{\partial\gamma} + \frac{\partial^2 c}{\partial\gamma\partial\gamma'}M_1\right) + M_1 M_2 \frac{\partial c}{\partial\gamma} \frac{\partial c}{\partial\gamma'} \left(h^2 + \frac{\partial h}{\partial c}\right), \label{eq:jpdf}
\end{align}
where,
\begin{align}
h =& \frac{\ln M_2}{c} - B c \Bigg[\frac{\beta\tilde{\mu}\sqrt{\tilde{a}}}{c^2 + \tilde{a}} + \frac{(1-\beta)\tilde{\mu}\alpha\sqrt{\tilde{a}}}{c^2 \alpha + \tilde{a}}+\frac{(1-\tilde{\mu})\sqrt{1-\tilde{a}}}{c^2 + 1-\tilde{a}}\Bigg],\\
\frac{\partial M_1}{\partial\gamma} = & -\frac{\tilde{a}(1-\tilde{a})}{(1+\tilde{a}\gamma-\tilde{a})^2}, \\
\frac{\partial M_1}{\partial\gamma'} = & -\frac{\tilde{a}(1-\tilde{a})}{[\gamma'(1-\tilde{a})+\tilde{a}]^2}, \\
\frac{\partial c}{\partial\gamma} = & \frac{1}{2\gamma(1-\gamma\gamma')} \left( c - \frac{\gamma'(1-\tilde{a})}{c} \right), \\
\frac{\partial c}{\partial\gamma'} = & \frac{1}{2\gamma'(1-\gamma\gamma')} \left( c - \frac{\gamma\tilde{a}}{c} \right), \\
\frac{\partial^2 c}{\partial\gamma\partial\gamma'} = & \frac{1}{4(1-\gamma\gamma')^2} \Bigg[3c + \frac{1}{c} - \frac{\tilde{a}(1-\tilde{a})}{c^3}\Bigg],\\
\frac{\partial h}{\partial c} = & - 2B \Bigg[\frac{\beta\tilde{\mu}\tilde{a}^{3/2}}{(c^2 + \tilde{a})^2} + \frac{(1-\beta)\tilde{\mu}\tilde{a}^{3/2}\alpha}{(c^2 \alpha + \tilde{a})^2}+\frac{(1-\tilde{\mu})(1-\tilde{a})^{3/2}}{(c^2 + 1-\tilde{a})^2}\Bigg]. \label{dh_dc}
\end{align}

\section{List of abbreviations.}
\begin{table}[H]
\label{tab:ListOfAbbr}
\center
\begin{tabular}{|l|l|}
\hline
{\bf Abbreviation} & {\bf Description}\\ \hline
HetNet & Heterogeneous Network\\ \hline
PPP & Poisson Point Process\\ \hline
BS & Base Station\\ \hline
MBS & Macro Base Station\\ \hline
PBS & Pico Base Station\\ \hline
MOI & Macrocell of Interest\\ \hline
POI & Picocell of Interest\\ \hline
UE & User Equipment\\ \hline
MUE & Macro User Equipment\\ \hline
PUE & Pico User Equipment\\ \hline
USF & Uncoordinated Subframe\\ \hline
CSF & Coordinated Subframe\\ \hline
eICIC & Enhanced Inter-cell Interference Coordination\\ \hline
FeICIC & Further Enhanced Inter-cell Interference Coordination\\ \hline
REB & Range Expansion Bias\\ \hline
SE & Spectral Efficiency\\ \hline
\end{tabular}
\end{table}

\end{appendices}

\bibliographystyle{IEEEtran}
\bibliography{Arxiv_article}

\begin{thebibliography}{10}
\providecommand{\url}[1]{#1}
\csname url@samestyle\endcsname
\providecommand{\newblock}{\relax}
\providecommand{\bibinfo}[2]{#2}
\providecommand{\BIBentrySTDinterwordspacing}{\spaceskip=0pt\relax}
\providecommand{\BIBentryALTinterwordstretchfactor}{4}
\providecommand{\BIBentryALTinterwordspacing}{\spaceskip=\fontdimen2\font plus
\BIBentryALTinterwordstretchfactor\fontdimen3\font minus
  \fontdimen4\font\relax}
\providecommand{\BIBforeignlanguage}[2]{{%
\expandafter\ifx\csname l@#1\endcsname\relax
\typeout{** WARNING: IEEEtran.bst: No hyphenation pattern has been}%
\typeout{** loaded for the language `#1'. Using the pattern for}%
\typeout{** the default language instead.}%
\else
\language=\csname l@#1\endcsname
\fi
#2}}
\providecommand{\BIBdecl}{\relax}
\BIBdecl

\bibitem{6524460}
H.~Elsawy, E.~Hossain, and M.~Haenggi, ``Stochastic geometry for modeling,
  analysis, and design of multi-tier and cognitive cellular wireless networks:
  A survey,'' \emph{Communications Surveys Tutorials, IEEE}, vol.~15, no.~3,
  pp. 996--1019, Third 2013.

\bibitem{6042301}
J.~Andrews, F.~Baccelli, and R.~Ganti, ``A tractable approach to coverage and
  rate in cellular networks,'' \emph{IEEE Transactions on Communications},
  vol.~59, no.~11, pp. 3122--3134, Nov. 2011.

\bibitem{5962727}
R.~Ganti, F.~Baccelli, and J.~Andrews, ``A new way of computing rate in
  cellular networks,'' in \emph{Proceedings of the IEEE Int. Conf. Commun.
  (ICC)}, Kyoto, Japan, June 2011, pp. 1--5.

\bibitem{5621983}
J.~Andrews, R.~Ganti, M.~Haenggi, N.~Jindal, and S.~Weber, ``A primer on
  spatial modeling and analysis in wireless networks,'' \emph{Communications
  Magazine, IEEE}, vol.~48, no.~11, pp. 156--163, November 2010.

\bibitem{5934671}
M.~Haenggi, ``Mean interference in hard-core wireless networks,''
  \emph{Communications Letters, IEEE}, vol.~15, no.~8, pp. 792--794, August
  2011.

\bibitem{5226957}
M.~Haenggi, J.~Andrews, F.~Baccelli, O.~Dousse, and M.~Franceschetti,
  ``Stochastic geometry and random graphs for the analysis and design of
  wireless networks,'' \emph{Selected Areas in Communications, IEEE Journal
  on}, vol.~27, no.~7, pp. 1029--1046, September 2009.

\bibitem{sayan-icc-ws}
S.~Mukherjee, ``Distribution of downlink {SINR} in heterogeneous cellular
  networks,'' \emph{{IEEE J. Select. Areas Commun. (JSAC), Special Issue on
  Femtocell Networks}}, vol.~30, no.~3, pp. 575--585, Apr. 2012.

\bibitem{Robert_TSP_2012}
R.~Heath, M.~Kountouris, and T.~Bai, ``Modeling heterogeneous network
  interference using poisson point processes,'' \emph{IEEE Trans. Signal
  Process.}, vol.~61, no.~16, pp. 4114--4126, Aug. 2013.

\bibitem{Andrews_Globecom_2011_ICIC}
T.~Novlan, R.~Ganti, and J.~Andrews, ``Coverage in two-tier cellular networks
  with fractional frequency reuse,'' in \emph{Proceedings of the IEEE Global
  Telecommun. Conf. (GLOBECOM)}, Houston, TX, Dec. 2011, pp. 1--5.

\bibitem{dhillon-ita}
H.~S. Dhillon, R.~K. Ganti, F.~Baccelli, and J.~G. Andrews, ``Modeling and
  analysis of {K}-tier downlink heterogeneous cellular networks,'' \emph{{IEEE
  J. Select. Areas Commun. (JSAC), Special Issue on Femtocell Networks}},
  vol.~30, no.~3, pp. 550--560, Apr. 2012.

\bibitem{Sayan_ICC_2012}
S.~Mukherjee, ``Downlink {SINR} distribution in a heterogeneous cellular
  wireless network with biased cell association,'' in \emph{Proceedings of the
  IEEE Int. Conf. Commun. (ICC)}, Ottawa, Canada, June 2012, pp. 6780 --6786.

\bibitem{Andrews_Globecom2011_RE}
H.-S. Jo, Y.~J. Sang, P.~Xia, and J.~Andrews, ``Outage probability for
  heterogeneous cellular networks with biased cell association,'' in
  \emph{Proceedings of the IEEE Global Telecommun. Conf. (GLOBECOM)}, Houston,
  TX, Dec. 2011, pp. 1--5.

\bibitem{RealBS_Lee}
\BIBentryALTinterwordspacing
C.-H. Lee, C.-Y. Shih, and Y.-S. Chen,
  ``\BIBforeignlanguage{English}{Stochastic geometry based models for modeling
  cellular networks in urban areas},''
  \emph{\BIBforeignlanguage{English}{Springer Wireless Networks}}, vol.~19,
  no.~6, pp. 1063--1072, 2013. [Online]. Available:
  \url{http://dx.doi.org/10.1007/s11276-012-0518-0}
\BIBentrySTDinterwordspacing

\bibitem{opencellid}
\BIBentryALTinterwordspacing
{OpenCellID} website. [Online]. Available: \url{www.opencellid.org}
\BIBentrySTDinterwordspacing

\bibitem{Lopez_perez2011HetNet}
D.~L\'opez-P\'erez, I.~G\"uven\c{c}, G.~de~la Roche, M.~Kountouris, T.~Q. Quek,
  and J.~Zhang, ``Enhanced inter-cell interference coordination challenges in
  heterogeneous networks,'' \emph{{IEEE Wireless Commun. Mag.}}, vol.~18,
  no.~3, pp. 22--31, June 2011.

\bibitem{Jeff_2013_arxiv}
S.~Singh and J.~G. Andrews, ``Joint resource partitioning and offloading in
  heterogeneous cellular networks,'' \emph{CoRR}, vol. abs/1303.7039, 2013.

\bibitem{Jeff_IEEE_Trans_2013}
S.~Singh, H.~S. Dhillon, and J.~G. Andrews, ``Offloading in heterogeneous
  networks: Modeling, analysis, and design insights,'' \emph{IEEE Trans.
  Wireless Commun.}, vol.~12, no.~5, pp. 2484--2497, May 2013.

\bibitem{Asilomar_2011}
S.~Mukherjee and I.~Guvenc, ``Effects of range expansion and interference
  coordination on capacity and fairness in heterogeneous networks,'' in
  \emph{Proceedings of the IEEE Asilomar Conf. Sig., Syst., Computers}, vol.~1,
  Monterey, CA, Nov. 2011, pp. 1855--1859.

\bibitem{CapAnalysis_Globecom_2013}
A.~Merwaday, S.~Mukherjee, and I.~Guvenc, ``On the capacity analysis of hetnets
  with range expansion and {eICIC},'' in \emph{Proceedings of the IEEE Global
  Commun. Conf. (GLOBECOM)}, Atlanta, GA, Dec 2013.

\bibitem{R1-113806_Panasonic}
Panasonic, ``Performance study on {ABS} with reduced macro power,'' 3GPP
  TSG-RAN WG1, Tech. Rep. R1-113806, Nov. 2011.

\bibitem{Morimoto_IEICE_2013}
A.~Morimoto, N.~Miki, and Y.~Okumura, ``Investigation of inter-cell
  interference coordination applying transmission power reduction in
  heterogeneous networks for {LTE}-advanced downlink,'' \emph{IEICE Trans. on
  Commun.}, vol. E96-B, no.~6, pp. 1327--1337, June 2013.

\bibitem{6289196}
M.~Al-Rawi, J.~Huschke, and M.~Sedra, ``Dynamic protected-subframe density
  configuration in {LTE} heterogeneous networks,'' in \emph{Proceedings of the
  IEEE Int. Conf. Computer Commun. Net. (ICCCN)}, Munich, Germany, 2012, pp.
  1--6.

\bibitem{LTE_Rel_11_Overview}
\BIBentryALTinterwordspacing
``Overview of 3gpp release 11 v0.1.7,'' Dec. 2013. [Online]. Available:
  \url{http://www.3gpp.org/ftp/Information/WORK\_PLAN/Description\_Releases/Rel-11\_description\_20131224.zip}
\BIBentrySTDinterwordspacing

\bibitem{MpactWebsite}
\BIBentryALTinterwordspacing
Mpact lab data management. [Online]. Available:
  \url{http://www.mpact.fiu.edu/data-management/}
\BIBentrySTDinterwordspacing

\bibitem{SiteFinder}
\BIBentryALTinterwordspacing
Sitefinder website. [Online]. Available:
  \url{http://www.sitefinder.ofcom.org.uk}
\BIBentrySTDinterwordspacing

\bibitem{Sayan_ICC_2011}
S.~Mukherjee, ``{UE} coverage in {LTE} macro network with mixed {CSG} and open
  access femto overlay,'' in \emph{Proceedings of the IEEE Int. Conf. Commun.
  (ICC) Workshops}, Kyoto, Japan, June 2011, pp. 1--6.

\bibitem{Yu_2012_TVT}
S.~M. Yu and S.-L. Kim, ``Downlink capacity and base station density in
  cellular networks,'' in \emph{Proceedings of the IEEE SpaSWiN workshop (in
  conjunction with WiOpt)}, Tsukuba Science City, Japan, May 2013, pp.
  119--124.

\bibitem{1003822}
P.~Viswanath, D.~Tse, and R.~Laroia, ``Opportunistic beamforming using dumb
  antennas,'' \emph{IEEE Trans. Inf. Theory}, vol.~48, no.~6, pp. 1277--1294,
  Aug. 2002.

\bibitem{6362834}
M.~R. Jeong and N.~Miki, ``A comparative study on scheduling restriction
  schemes for lte-advanced networks,'' in \emph{Proceedings of the IEEE 23rd
  Int. Symp. Personal Indoor and Mobile Radio Communications (PIMRC)}, Sydney,
  Australia, Sept. 2012, pp. 488--495.

\end{thebibliography}

\end{document}